\documentclass[sigplan,screen]{acmart}

\usepackage[normalem]{ulem}
\usepackage{amsmath,amsthm}
\usepackage{cleveref}
\usepackage{listings, fancyvrb, multicol}
\usepackage{graphicx}
\renewcommand\thepage{}

\makeatletter
\lst@Key{countblanklines}{true}[t]%
{\lstKV@SetIf{#1}\lst@ifcountblanklines}

\lst@AddToHook{OnEmptyLine}{%
	\lst@ifnumberblanklines\else%
	\lst@ifcountblanklines\else%
	\advance\c@lstnumber-\@ne\relax%
	\fi%
	\fi}
\makeatother

\newcommand{\Changed}[1]{\noindent
	{\color{black}{#1}}}

\newcommand{\Michal}[1]{\noindent 
  {\color{black}{#1}}}

\newcommand{\he}[1]{\noindent
	{#1}}

\newtheorem{theorem}{Theorem}[section]
\newtheorem{lemma}[theorem]{Lemma}

\newtheorem{claim}[theorem]{Claim}

\newtheorem{definition}{Definition}

\newtheorem{property}{Property}
\newtheorem{protocol}{Protocol}
\newtheorem{supplement}{Supplement}

\setcopyright{none}

\newif\ifflushmark
\flushmarkfalse

\newif\ifonenode

\newif\iforiginalparent
\originalparenttrue

\newcommand{\valueChange}{valueChange}

\newcommand{\makePersist}{makePersistent}
\newcommand{\flush}{flush}
\newcommand{\fence}{fence}

\newcommand{\traverse}{traverse}
\newcommand{\critical}{critical}
\newcommand{\reach}{ensureReachable}

\newcommand{\ourDS}{traversal data structure}
\newcommand{\OurDS}{Traversal data structure}
\newcommand{\durableOurDS}{NVTraverse data structure}
\newcommand{\DurableOurDS}{NVTraverse Data Structure}

\newcommand{\disconnect}{disconnect}

\newcommand{\persist}{persist}

\newcommand{\modret}{externally visible}
\newcommand{\findentry}{findEntry}

\renewcommand\footnotetextcopyrightpermission[1]{}

\author{Michal Friedman}
\affiliation{Technion, Israel}       
\email{michal.f@cs.technion.ac.il}
\authornote{The first three authors contributed equally to this work. Their order of appearance in the paper was selected randomly.}

\author{Naama Ben-David}
\affiliation{CMU, USA}
\email{nbendavi@cs.cmu.edu}
\authornotemark[1]

\author{Yuanhao Wei}
\affiliation{CMU, USA}        
\email{yuanhao1@cs.cmu.edu}
\authornotemark[1]

\author{Guy E. Blelloch}
\affiliation{CMU, USA}         
\email{guyb@cs.cmu.edu}

\author{Erez Petrank}
\affiliation{Technion, Israel}     
\email{erez@cs.technion.ac.il}

\title[NVTraverse]{NVTraverse: In NVRAM Data Structures, the Destination is More Important than the Journey}



\setcopyright{acmcopyright}
\acmPrice{15.00}
\acmDOI{10.1145/3385412.3386031}
\acmYear{2020}
\copyrightyear{2020}
\acmSubmissionID{pldi20main-p822-p}
\acmISBN{978-1-4503-7613-6/20/06}
\acmConference[PLDI '20]{Proceedings of the 41st ACM SIGPLAN International Conference on Programming Language Design and Implementation}{June 15--20, 2020}{London, UK}
\acmBooktitle{Proceedings of the 41st ACM SIGPLAN International Conference on Programming Language Design and Implementation (PLDI '20), June 15--20, 2020, London, UK}

\begin{document}
	
	\date{}
			\begin{abstract}
		The recent availability of fast, dense, byte-addressable non-volatile memory has led to increasing interest in the problem of designing and specifying durable
		data structures that can recover from system crashes. However,
		designing durable concurrent data structures that are
		efficient and also satisfy a correctness criterion has proven
		to be very difficult, leading many algorithms to be
		inefficient or incorrect in a concurrent setting. In this
		paper, we present a general transformation that
		takes a lock-free data structure from a general class called
		\emph{\ourDS} (that we formally define) and automatically
		transforms it into an implementation of the data structure for
		the NVRAM setting \he{that is provably durably linearizable and highly efficient.} The transformation hinges on the observation
		that many data structure operations begin with a traversal
		phase that does not need to be persisted, and thus we only
		begin persisting when the traversal reaches its destination.
		We demonstrate the transformation's efficiency through
		extensive measurements on a system with Intel's recently
		released Optane DC persistent memory, showing that it can outperform competitors on many workloads.
	\end{abstract}
	\maketitle
	\pagestyle{plain}


	
	\section{Introduction}
Now that non-volatile random access memory (NVRAM) has finally hit the market, the question of how to best make use of it is more pressing than ever.
NVRAM offers byte-addressable persistent memory at speeds comparable with DRAM. This memory technology now can co-exist with DRAM on the newest Intel machines, and may largely replace DRAM in the future.  
Upon a system crash in a machine with NVRAM, data stored in main memory will not be lost. However, without further technological advancements, caches and registers are expected to remain volatile, losing their contents upon a crash. Thus, NVRAM yields a new model and opportunity for programs running on such machines---how can we take advantage of persistent main memory to recover a program after a crash, despite losing values in cache and registers?

\Changed{
One challenge of using NVRAM is that a system crash may occur part way through a large update, leaving the memory with some, but not all, of the changes that should have been executed. In some concurrent lock-based programs, the state of memory after a partial update may not be consistent, and may be unsafe for other processes to observe. Furthermore, without knowing the entire update operation and what changes it should have made, it may be impossible to return the memory to a consistent state. This may require heavy-duty mechanisms, like logging or copying, to be employed.

Interestingly, a well-studied class of programs called \emph{lock-free algorithms} ensures that the memory is always in a consistent state, even during long updates. In a nutshell, lock-freedom requires that processes be able to execute operations on the shared state regardless of the slow progress of other processes in the system. This means that even if a process is swapped-out part way through its update, other processes can continue their execution. Thus, lock-free algorithms are a very natural fit for use in NVRAM. }

However, using NVRAM introduces still more challenges. Because of their small size, caches inherently require evicting cache lines back to main memory. On modern caches, these evictions are performed automatically when needed, thus offering a fast, transparent interface for their user. Yet, when main memory is persistent, this can create complications; it is possible that values written to cache later in a program get evicted earlier than others, thus making main memory hold out-of-order values. When a system crash occurs, this can leave the memory in an inconsistent state, from which it may not be possible to recover.
This problem is especially challenging when multiple processes access the same memory locations.
\Changed{Lock-free algorithms are generally not designed to handle such reordering of memory updates.}

To fix this, explicit flushes and fences can be introduced into programs to force certain changes to appear in main memory before others. In particular, if we add a flush and a fence instruction between every two synchronized instructions of a process, 
%
main memory remains consistent. \citet{izraelevitz2016linearizability} formalized this intuition and showed that this technique indeed leads to a consistent memory state in all lock-free programs, regardless of crashes. Thus, many theoretical papers have focused on a model in which changes to the memory are assumed to be persisted immediately, and always in the order they occur~\citep{bendavid2019delay,blelloch2018parallel,berryhill2016robust,attiya2018nesting}.
However, fences are notoriously expensive, causing this approach to be prohibitively slow, despite guaranteeing correctness.

A lot of research has instead focused on 
decreasing the amount of flushing needed during regular execution to be able to recover~\citep{shull2019autopersist,chauhan2016nvmove,chen2015persistent,liu2017dudetm,lee2019recipe}.
 However, the notion of "being able to recover" is flexible: do we allow the loss of some progress made before the crash?  Without defining this clearly, unexpected behavior can result from algorithms that are seemingly `correct'. 
 
Significant work has also focused on defining these goals~\citep{aguilera2003strict,guerraoui2004robust,chen2015persistent}. 
\citet{izraelevitz2016linearizability} introduced the notion of \emph{durable linearizability}.
In a nutshell, a concurrent data structure is said to be durably linearizable if all executions on it are linearizable once crash events are removed. This disallows the state of the data structure in memory to be corrupted by a crash, and does not allow the effect of completed operations to be lost.
In their paper, Izraelevitz et al. posit that durable linearizability may be prohibitively expensive. 

Since then, several works presented hand-tuned algorithms that achieve durable linearizability while performing reasonably well~\citep{friedman2018persistent,david2018logfree}. However, they only show specific data structures, and do not yield a general way of designing practical durably linearizable data structures. 
The difficulty of this task can be traced back to understanding the \emph{dependencies} between operations in the algorithm.
Only expertise about persistence, combined with a deep understanding of an algorithm and careful reasoning about its inherent dependencies, has so far been productive in finding correct and more efficient solutions.

\Michal{
A different line of work introduced persistent transactions \cite{volos2011mnemosyne, correia2018romulus, rcaec19onefile, coburn2011nvheaps, pmdk,marathe2018persistent,zardoshti2019optimizing,kolli2016high, Wu2019ArchitectureAwareHP}. Persistent transactions provide an easier model for programming, because the transaction either persists as a whole in the non-volatile memory or does not take an effect. This allows executing several operations atomically. However, as in the non-persistent case, the use of transactions brings ease of programming at the cost of lower performance, as shown in \Cref{sec:exp}. In this work we focus on issuing single operations persistently on lock-free data structures. Lock-free transactional data structures~\cite{elizarov19} allow executing several operations atomically on lock-free data structures. An interesting open question is how to make these transactions persistent, but this is outside the scope of this work. }

\textbf{Our contributions.}
In this paper, we provide a technique that achieves the best of both worlds for a large class of lock-free data structure implementations; \Changed{we show an automatic transformation that can be applied to lock-free algorithms of a certain form, and makes such data structures persistent and efficient, with well-defined correctness guarantees that are provably correct.}
We take a substantial step in removing the need for expert familiarity with an algorithm to make an efficient durably linearizable version of it.
Our key insight is that many lock-free concurrent data structure implementations begin operations with a \emph{traversal} of the data structure, and that, intuitively, the values read along this traversal do not affect the operation's behavior after the traversal.

%

We formalize a large class of lock-free linearizable algorithms we call \emph{\ourDS{s}}. \OurDS{s} are node-based tree data structures whose operations first traverse the data structure, and then perform modifications on nodes that descend from where their traversal stopped. The traversal is guaranteed to only make local decisions at every point in time, not relying on previous nodes to determine how to proceed from the current one. These algorithms also follow some natural rules when removing nodes from the data structure.

We show that many existing pragmatic concurrent algorithms can easily be converted into \ourDS{s}, without losing their efficiency. We use Harris's linked-list~\citep{harris2001pragmatic} as a running example throughout the paper to help with exposition. 
Other data structures implementations, like common BST, (a,b)-tree, and hash table algorithms~\cite{ellen2010non,brown2017template,brown2014general}, can also be easily converted.  Furthermore, \ourDS{s} capture not just set data structures, but also queues, stacks, priority queues, skiplists, augmented trees, and others. 
Most requirements of \ourDS{s} are naturally satisfied by many lock-free data structures.
Thus, it is not hard to transform a new data structure implementation into a \ourDS.

After defining \ourDS{s}, we show how to automatically inject flush and fence instructions into a \ourDS{} to make it durably linearizable. The key benefit of our approach is that no flushes are needed during most of the traversal. Of all memory locations that are read, only a few at the end of the traversal must be flushed. Because of the careful way in which the traversal is defined, these can be automatically identified.
After the traversal, all further locations that the operation accesses must be flushed. However, in most operations, the traversal encapsulates a large majority of the work. 

We formally prove that the flushes and fences that we specify are sufficient for durability for all \ourDS{s}. Thus, this paper presents the first practical, provably correct implementation of many durable data structures; the only previously known durable algorithm that was proven correct is the DurableQueue of~\citep{friedman2018persistent}.

Finally, we experimentally evaluate the algorithms that result from our transformation, by transforming a list, two binary search trees, a skiplist, and a hash table to the traversal form and then injecting flushes automatically to obtain a durable data structure. We compare our implementations to those that result from the general transformation of \citet{izraelevitz2016linearizability}, the original (non-persistent) version of each algorithm, the OneFile transactional memory~\cite{rcaec19onefile}, and the hand-tuned durable version presented by \citet{david2018logfree}. 
We reclaim objects using simple epoch-based memory management.
Our results show that persisted \ourDS{s}, or \emph{\durableOurDS{s}}, outperform \citet{izraelevitz2016linearizability}'s construction significantly on all workloads. Thus, \durableOurDS{s} are a much better alternative as a general transformation for concurrent data structures to become durable.
Furthermore, \durableOurDS{s} outperform \citet{david2018logfree}'s data structures on about half of the workloads; those with lower thread counts or larger data structure sizes. This provides some interesting insights on the tradeoffs between flushes, fences, and writes when it comes to contention.
%

\Changed{
Using the method proposed in this work involves two steps. The first step (which is manual) involves making sure that the target lock-free data structure is in the traversal format. The second step (which is automatic) involves adding flushes and fences to make the lock-free data structure durable. The second step is the major contribution of this work, because it spares programmers the effort of reasoning about persistence. On the other hand, the definition of \ourDS{s} is not as simple as we would have wanted it to be. While many data structures are already in traversal form, the programmer must verify that this is the case for their data structure before using our transformation. Sometimes, small modifications are required to make a data structure a traversal one. Simplifying the definition of traversal data structures (while keeping the correctness and generality of the transformation) is an interesting open problem. We stress that the contribution of this paper is significant, as the alternative approaches available today for building durable linearizable data structures are either (1) to use the simple but inefficient transformation of Izraelevitz et al. \cite{izraelevitz2016linearizability}, or (2) to carefully reason about crash resilience to determine where flushes and fences should be inserted for each new data structure. 

}

In summary, our contributions are as follows.

\begin{itemize}
	\item We define a large class of concurrent algorithms called \emph{\ourDS{s}}. Many known lock-free data structures can be easily \he{put in} traversal form.
	\item We show how to automatically transform any \ourDS{} to become durable with significantly fewer flushes and fences than previously known general techniques.
	\item We prove that our construction is correct.
	\item We implement several data structures using our transformation and evaluate their performance compared to state-of-the-art constructions.
	\vspace{-0.2cm}
\end{itemize}


	\section{Model and Preliminaries}


In this paper, we show how to convert a large class of algorithms designed for the standard shared-memory model into algorithms that maintain their correctness in non-volatile (persistent) memory with crashes. 
Thus, we first present a recap of the classic shared-memory model, and then discuss the changes that are added in our persistent memory model.
Throughout the paper, we sometimes say that a node $n$ in a tree data structure is \emph{above (resp. below)} another node $n'$ if $n$ is an ancestor (resp. descendant) of $n'$.

\textbf{Classic shared memory.}
We consider an asynchronous shared-memory system in which processes execute \emph{operations} on data structures. 
Data structure operations can be implemented using instructions local to each process, including \emph{return statements}, as well as shared atomic \emph{read}, \emph{write}, and \emph{compare-and-swap (CAS)} instructions.
We sometimes refer to write and successful CAS instructions collectively as \emph{modifying} instructions, and to modifying instructions, and return statements collectively as \emph{\modret} instructions.
\he{In the experiments section, we use the term \emph{threads} instead of \emph{processes}.}

\textbf{Linearizability and lock-freedom.}
We say that an execution history is \emph{linearizable}~\citep{herlihy1990linearizability} if every operation takes effect atomically at some point during its execution interval.
\Changed{A data structure is \emph{lock-free} if it guarantees that at least one process makes progress, if processes are run sufficiently long. This means that a slow/halted process may not block others, unlike when using locks.}

\textbf{Persistent memory.}
In the persistent memory model,
there are two levels of memory--- \emph{volatile} and \emph{persistent} memory, which roughly correspond to cache and NVRAM. 
All memory accesses (both local and shared) are to volatile memory. Values in volatile memory can be written back to persistent memory, or \emph{persisted}, in a few different ways; a value could be persisted \emph{implicitly} by the system, corresponding to an automatic cache eviction, or \emph{explicitly} by a process, by first executing a \emph{flush} instruction, followed by a \emph{fence}. 
We assume that when a fence is executed by a process $p$, all locations that were flushed by $p$ since $p$'s last fence instruction get \persist{ed}. 
We say that a value has been \emph{persisted} by time $t$ if the value reaches persistent memory by time $t$, regardless of whether it was done implicitly or explicitly.
Note that persisting is done on \emph{memory locations}. However, it is sometimes convenient to discuss \emph{modifications} to memory being \persist{ed}. We say that a modifying instruction $m$ on location $\ell$ is \persist{ed} if $\ell$ was \persist{ed} since $m$ was executed. A modifying instruction $m$ is said to be \emph{pending} if it has been executed but not persisted. 

\he{In the persistent memory model, \emph{crash}es may also occur.}
A crash event causes the state of volatile memory to be lost, but does not affect the state of persistent memory.
Thus, all modifications that were pending at the time of the crash are lost, but all others remain. 
Each data structure may have a \emph{recovery} operation in addition to its other operations. Processes call the recovery operation before any other operation after a crash event, and may not call the recovery operation at any other time.
\he{The recovery operation can be run concurrently with other operations on the data structure.}
We say that an execution history is \emph{durably linearizable}~\citep{izraelevitz2016linearizability} if, after removing all crash events, 
the resulting history is linearizable. \Changed{In particular, this means that the effect of completed operations may not be lost, and operations that were in progress at the time of a crash must either take effect completely, or leave no effect on the data structure. Furthermore, if an operation does take effect, then all the operations it depends on must also have taken effect.}
			\vspace{-0.2cm}

\subsection{Running Example: Harris's Linked List}

Throughout the paper, we refer to the linked-list presented by \citet{harris2001pragmatic} when discussing properties of \ourDS{s}. We now briefly describe how this algorithm works.

\citet{harris2001pragmatic} presented a pragmatic linearizable lock-free implementation of a sorted linked-list. The linked-list is based on nodes with an immutable \emph{key} field and a mutable \emph{next} field, and implements three high-level operations: insert, delete, and find, which all take a key $k$ as input.  
Each of these operations is implemented in two stages: first, the helper function \emph{search} is called with key $k$, and after it returns, changes to the data structure are made on the nodes that the search function returned. The search function always returns two adjacent nodes, \emph{left} and \emph{right}, where \emph{right} is the first element in the list whose key is greater than or equal to $k$, and \emph{left} is the node immediately before it.

To insert a node, the operation simply initializes a node with \emph{key} $k$ and \emph{next} pointing to the right node returned from the search function, and then swings \emph{left}'s \emph{next} pointer to point to the newly initialized node (using a CAS with expected value pointing to \emph{right}). If the CAS fails, the insert operation restarts.
The find operation is even simpler; if \emph{right}'s key is $k$, then it returns true, and otherwise it returns~false.

The subtlety comes in in the delete operation.
The \emph{next} pointer of each node has one bit reserved as a special \emph{mark} bit. If this bit is set, then this node is considered \emph{marked}, meaning that there is a pending delete operation trying to delete this node. If a node is marked, we say that it is \emph{logically deleted}.
More specifically, a delete operation, after calling the search function, uses a CAS to mark the right node returned by the search, if the key of \emph{right} is $k$. After successfully marking the right node, the delete operation then \emph{physically deletes} the right node by swinging \emph{left}'s \emph{next} pointer from \emph{right} to \emph{right.next}.
This two-step delete is crucial for correctness, avoiding synchronization problems that may arise when two concurrent list operations contend.

The search function guarantees that neither of the two nodes that it returns are marked, and that they are adjacent. To be able to guarantee this, the search function must help physically delete marked nodes.
Thus, the search function finds the \emph{right} node, which is the first unmarked node in the list whose key is greater than or equal to $k$, and the \emph{left} node, which is the last unmarked ancestor of \emph{right}.
Before returning, the search function physically deletes all nodes between the two nodes it intends to return.

\section{\OurDS{s}}
In this section, we introduce the class of data structures we call \emph{\ourDS{s}}, and the properties that all \ourDS{s} must satisfy.
In \Cref{sec:flushes}, we show an easy and efficient way to make any \ourDS{} durable. We begin with two simple yet important properties.

\begin{property}[Correctness]\label{prop:correct}
	A \ourDS{} is linearizable and lock-free.
\end{property}

\begin{property}[Core Tree]\label{prop:tree}
	A \ourDS{} is a node-based tree data structure. In addition to the tree, there may be other nodes and links that are auxiliary, and are only ever used as additional entry points into the tree. 
\end{property}


\he{The part of the data structure that needs to be persistent and survive a crash is called its \emph{core}. The other parts can be stored in volatile memory and recomputed following a crash. Property \ref{prop:tree} says that only the \emph{core} part of a \ourDS{} needs to be a tree}.

For example, a skiplist can be a \ourDS{}, since, while the entire structure is not a tree, only a linked list at the bottom level holds all the data in the skiplist, while the rest of the nodes and edges simply serve as a way to access the linked list faster. Similarly, data structures with several entry points, like a queue with a head and a tail, can be \ourDS{s} as well. Of course, all tree data structures fit this requirement.

\he{More precisely, the core of a \ourDS{} must be a \emph{down-tree}, meaning that all edges are directed and point away from the root. For simplicity, we use the term \emph{tree} in the rest of the paper.}

\Cref{prop:tree} is important since it simplifies reasoning about the data structure, and thus allows us to limit flush and fence instructions that need to be executed to make \ourDS{s} durable. Note that many pragmatic data structures, including queues, stacks, linked lists, BSTs, B+ trees, skiplists, and hash tables have a core-tree structure. Optionally, a \ourDS{} may also provide a function to reconstruct the structure around the core tree at any point in time. 
However, our persistent transformation maintains the correctness of the core tree regardless of whether such a function exists.

A \ourDS{} is composed of three methods: \emph{\findentry{}}, \emph{\traverse}, and \emph{\critical}. These are the only three methods through which a \ourDS{} may access shared memory, and they are always called in this order. The operation \findentry{} takes in the input of the operation, and outputs an entry point into the core tree. That is, the \findentry{} method is used to determine which shortcuts to take. This can be the head of a linked-list, a tail of a queue, or a node of the lowest level of the skip list, from which we traverse other lowest-level nodes. Note that \findentry{}, and is allowed to simply return the root of the tree data structure, e.g., the head of a linked-list.

Once an entry point is identified, a \ourDS{} operation starts a \traverse{} from that point, at the end of which it moves to the critical part, in which it may make changes to the data structure, or determine the operation's return value.
%
The \critical{} method may also determine that the operation must \emph{restart} with the same input values as before. However, the \traverse{} method may not modify the shared state at all.
The operation execution between the beginning of the \findentry{} method and the return or restart statement in the \critical{} method is called an \emph{operation attempt}.
Each operation attempt may only have one call to the \findentry{} method, followed by one call to the \traverse{} method, followed by one call to a \critical{} method.
The layout of an operation of a \ourDS{} is shown in Algorithm~\ref{alg:ourDS}.
Operations may not depend on information local to the process running them; an operation only has access to data provided in its arguments, one of which is the root of the data structure. The operation may traverse shared memory, so it can read anything accessible in shared memory from the root. No other argument is a pointer to shared memory. This requirement is formalized in \Cref{prop:procIndependent}. 

\begin{property}[Operation Data]\label{prop:procIndependent}
	Each operation attempt only has access to its input arguments, of which the root of the data structure is the only pointer to shared memory. Furthermore, it accesses the shared data only through the layout outlined in Algorithm~\ref{alg:ourDS}.
\end{property}

By similarity to the original algorithm of Harris, the traversal version of Harris's linked-list is linearizable and lock-free (thereby satisfying \Cref{prop:correct}), and the data structure is a tree (thereby satisfying \Cref{prop:tree}). Furthermore, Harris's linked-list implementation gets the root of the list as the only entry point to the data structure and it only uses the input  arguments in each operation attempt. \Changed{Each operation of Harris's linked-list can be easily modified to only use the three methods shown in \Cref{alg:ourDS}. The findEntry method simply returns the root. The traverse method encompasses the \emph{search} portion of the operation, but does not physically delete any marked nodes. Instead, the \traverse{} method returns the \emph{left} and \emph{right} nodes identified, as well as any marked nodes between them. The rest of the operation is executed in the critical method.} Therefore, Harris's linked-list can easily be converted to satisfy \Cref{prop:procIndependent}.

In the rest of this section, we discuss further requirements on \ourDS{s}, which fall under two categories: traversal and disconnection behavior (when deleting a node). 
We also show that Harris's linked-list can easily be made to satisfy these properties, and thus can be converted into a \ourDS. 

From now on, when we refer to a \ourDS{}, we mean only its core tree, unless otherwise specified.

	\renewcommand{\figurename}{Algorithm}
\begin{figure}
	\caption{Operation in a \ourDS}
	\begin{lstlisting}
@$\textbf{T}$@  operation (@$\textbf{Node}$@  root, @$\textbf{T'}$@ input) {
	while (true) {
		@$\textbf{Node}$@ entry = @\textbf{\findentry}@ (root, input);
		List<@$\textbf{Node}$@> nodes = @\textbf{\traverse}@ (entry, input);
		bool restart, @$\textbf{T}$@ val = @\textbf{\critical}@ (nodes, input);
		if (!restart)
			return val;	
	}	
}
\end{lstlisting}
\label{alg:ourDS}
\end{figure}

\subsection{Traversal}

Intuitively, we require the \traverse{} method to "behave like a traversal". It may only read shared data, but never modify it (\Cref{item:readTraversal} of \Cref{prop:traverse}), and may only use the data it reads to make a local decision on how to proceed.
The \traverse{} method starts at a given node, and has a \emph{stopping condition}. After it stops, it returns a suffix of the path that it read. In most cases the nodes that it returns are a very small subset of the ones it traversed. 
The most common use case of this is that the traversal stops once it finds a node with a certain key that it was looking for, and returns that node. However, we do not specify what this stopping condition is, or how many nodes are returned, to retain maximum flexibility. 


\Cref{item:stopTraversal} and \Cref{item:directionTraversal} of \Cref{prop:traverse} formalize the intuition that a traversal does not depend on everything it read, but only on the local node's information. The \traverse{} method proceeds through nodes one at a time, deciding whether to stop at the current node, using only fields of that node, and, if not, which child pointer to follow. The child pointer decision is made only based on immutable values of this node; intuitively, if a node has a immutable key and a mutable value then keys can be compared, but the node's value cannot be used.
\he{The stopping condition can make use of both mutable and immutable fields of the current node.}

If the \traverse{} method does stop at the current node, it then determines which nodes to return. This decision may only depend on the nodes returned; no information from earlier in the traversal can be taken into account (\Cref{item:returnTraversal} of \Cref{prop:traverse}). Intuitively, we allow the \traverse{} method to return multiple nodes since some lock-free data structures make changes on a \emph{neighborhood} of the node that their operation ultimately modifies.
%
Examples of such a data structure include Harris's linked-list~\citep{harris2001pragmatic}, Brown's general tree construction~\cite{brown2014general}, Ellen et al.'s BST~\cite{ellen2010non}, Herlihy et al.'s skip-list~\cite{herlihy2007simple}, and others. All of these data structures make changes on the parent or grandparent of their desired node, or find the most recent unmarked node under some marking mechanism.

Note that we allow arbitrary mutable values to be stored on each node. 
However, we add one more requirement. 
\Changed{Intuitively, a non-pointer-swing change on a node may not make a traversal return a later node than it would have had it not seen this change. More precisely, 
suppose traversal $T_1$ reads a non-pointer value $v$ on node $n$ and decides to stop at $n$. Consider another traversal $T_2$ with the same input as $T_1$ that
reads the same field after the value $v$ was modified. We require that $T_2$'s returned nodes be at or above $n$. 
Note that we consider a `marking' of a node to be a non-pointer value
modification, even though some algorithms place the mark physically on the pointer field.
It is easiest to understand this requirement by thinking about deletion marks; suppose $v$ is a mark bit, and node $n$ is marked for deletion between $T_1$ and $T_2$. Then if $T_1$'s search stopped at $n$ (i.e., it was looking for the key stored at $n$ and found it), it's possible $T_2$ may continue further, since $n$ is now `deleted'. However, when $T_2$ returns, it must return a node above $n$, since the operation that called it must be able to conclude that $n$'s key has been deleted. This will be important for persisting changes that affected the return value of $T_2$'s operation.
}  
%
This property can be thought of as a \emph{stability} property of the traversal; it may stop earlier, but may not be arbitrarily perturbed by changes on its way. We formalize this in \Cref{item:stableTraversal} of \Cref{prop:traverse}.
We say that a \emph{\valueChange{}} is any node modification of a non-pointer value (i.e., not a disconnection or an insertion).

\begin{property}[Traversal Behavior]\label{prop:traverse}

	The \traverse{} method must satisfy the following properties.
	\begin{enumerate}
		\item \label{item:readTraversal} \textbf{No Modification:} It does not modify shared memory.

		\item \label{item:stopTraversal} \textbf{Stopping Condition:} Only the current node is used to decide whether or not to continue traversing.

		\item  \label{item:directionTraversal} \textbf{Traversal Route:} Only \emph{immutable} values of the current node are used to determine which pointer of the current node to follow next.
		
		\item \label{item:returnTraversal} \textbf{Traversal Output:} The output may be any suffix of the path traversed. The decision of which nodes to return may only depend on data in those returned nodes.
		
		\item \label{item:stableTraversal} \textbf{Traversal Stability:} Consider two traversals $T_1$ and $T_2$ such that both of them have the same input and read the same field $f$ of the same node $n$. Let $m$ be a \emph{\valueChange{}} of $f$ that occurs after $T_1$'s read but before $T_2$'s read. If $T_1$ stopped at $n$, then $T_2$ returned $n$ or a node above $n$.

	\end{enumerate}
\end{property}

We now briefly show how the traversal version of Harris's linked list algorithm can satisfy \Cref{prop:traverse}.
Recall that the traversal of each operation in Harris's linked-list is inside the search function, which begins by finding the first node in the list that is unmarked and whose key is greater than or equal to the search's key input (this node is called the \emph{right} node). It then finds the closest preceding unmarked node, called the \emph{left} node. We define the search function up to \he{the right node} as the \traverse{} method. The \traverse{} method then returns all nodes from \emph{left} to \emph{right}. 
At every point along the traversal, it uses only fields of the current node to decide whether or not to stop. If not, it always reads the \emph{next} field and follows that pointer; its decision of how to continue its traversal does not depend on any mutable value that it reads.
The returned nodes depend only on values between \emph{left} and \emph{right}. 
Thus, Items \ref{item:readTraversal}, \ref{item:stopTraversal}, \ref{item:directionTraversal}, and \ref{item:returnTraversal} of \Cref{prop:traverse} are satisfied.

To see that \Cref{item:stableTraversal} is satisfied, note that the only \valueChange{}  in Harris's algorithm is the marking of a node for deletion. So, if a traversal $T_1$ stops at a node $n$ (i.e., $n$ is the \emph{right} node of the search), and a traversal $T_2$ with the same input sees $n$ marked, then $T_2$ would stop after $n$, but would return a node above $n$ ($T_2$'s \emph{left} and \emph{right} nodes must both be unmarked, and $n$ must be in between them).
%
%

\subsection{Critical Method: Node Disconnection}

The only restrictions we place on the \critical{} method's behavior are on how nodes are \emph{disconnected} from the data structure.
Disconnections may be executed to delete a node from the data structure, but some implementations may disconnect nodes to replace them with a more updated version, or to maintain some invariant about the structure of a tree.

Many lock-free data structure algorithms~\cite{harris2001pragmatic,brown2014general,ellen2010non,herlihy2007simple,aravind14bst} first logically delete nodes by \emph{marking} them for deletion before physically disconnecting them from the data structure. This technique prevents the logically deleted nodes from being further modified by any process, thus avoiding data loss upon their removal. We begin by defining marking. 

\begin{definition}[Mark]\label{def:mark}
	A node is \emph{marked} if the mark method has been called for it. Once a node is marked, no field in it can be modified.
\end{definition}

We require that before any node is disconnected, it is marked (Item \ref{item:markbeforedelete} of Property~\ref{prop:disconnect}). Sometimes, the marking of a node/nodes, denoted as $S$ consists of changing the fields in multiple nodes. 
For example, in the BST of Ellen \emph{et al.}~\cite{ellen2010non}, the parent of $S$'s and the grandparent of $S$'s values are changed to indicate the marking of $S$. They hold a descriptor indicating the disconnection that needs to be executed to remove $S$. Once the final field is written to specify the marking of $S$, the node is considered marked and cannot be modified.

Furthermore, marking is intended only for nodes to be removed from the data structure. To formalize that, \Cref{item:uniqueDisconnect} of Property~\ref{prop:disconnect} states that there is \emph{always} a legal instruction that can be executed to atomically disconnect a given marked, connected subset of nodes $S$ from a \ourDS. 
An instruction is considered \emph{legal} if it is performed in some extension of the current execution.
We further require that at each configuration, there is \emph{at most} one legal disconnect instruction for a contiguous set of marked nodes $S$.
This in effect means that the marks themselves must have enough information encoded in them to uniquely identify the disconnection instruction that may be executed. Some data structures, like Harris's linked list, achieve this trivially, since there is only ever one way to disconnect a node. Other data structures use \emph{operation descriptors} inside their marking protocol, which specify what deletion operation should be carried out~\cite{brown2014general,ellen2010non}.

Let $P^*(S)$ be all the nodes whose values are changed as part of the marking of $S$, indicating the unique physical disconnection that needs to be executed. We make an additional requirement for algorithms whose marking of a node includes changing values (marking) multiple nodes, including nodes that are not intended for deletion (as discussed following Definition~\ref{def:mark}). Denote all mark indications on nodes that are not intended for deletion as {\em external marks}. Such algorithms usually remove the external marks after the marked nodes have been disconnected. We specifically require that a thread that removes external marks either attempted to execute the disconnection, or has read the updated pointer whose update executed the disconnection. We do not allow a removal of external marks ``out of the blue'' by threads that
have not been involved in the disconnection or that have not seen the updated pointer. This property is formalized in Item~\ref{item:removeMark} below, and it holds for all known lock-free algorithm (that use extra marking indications outside the marked node).

It is also important that marked nodes can be removed from the data structure in any order. This property is formalized in \Cref{item:overlapDisconnect}.

\begin{property}[Disconnection Behavior]\label{prop:disconnect}
	In a \ourDS{}, node disconnections satisfy the following properties:
	\begin{enumerate}
		\item \label{item:markbeforedelete} \textbf{Mark Before Delete:} Before any node is disconnected from a \ourDS{}, it must be marked.
		\item \label{item:uniqueDisconnect} \textbf{Unique Disconnection:} 
		Consider a configuration $C$ and let $S$ be a connected subset of marked nodes in the core tree of a \ourDS{}. Let $P(S)$ be the parent of the root of $S$. If $P(S)$ is unmarked, then there is exactly one legal instruction on $P(S)$ that atomically disconnects exactly the nodes in $S$.
		\item \label{item:removeMark} \textbf{Delete Before Mark Removal:} Let $m$ be the unique disconnection instruction on $P(S)$ that atomically disconnects exactly the nodes in $S$, and let $l$ be the location of the pointer that is modified to disconnect $S$. Before a thread $t$ reverts an external mark, it must execute $m$ or fails to execute $m$ because another thread has done it earlier, or it must read $l$ after it was modified to disconnect $S$.
		\item \label{item:overlapDisconnect} \textbf{Irrelevant Disconnection Order:} Let $N =\{n_1 \ldots n_k\}$ be the set of nodes that were marked at configuration $C$. Let $E_1$ and $E_2$ be two executions that both start from the same configuration and only perform legal disconnecting operations. If all marked nodes are disconnected after $E_1$ and $E_2$, then the state of the nodes in the data structure after $E_1$ and after $E_2$ is the same.
	\end{enumerate}
\end{property}

We now argue that Harris's linked list satisfies Property~\ref{prop:disconnect}.
A node in Harris's linked list is considered marked if the lowest bit on its next pointer is set. Once this bit is set, the node becomes immutable. Item~\ref{item:markbeforedelete} of Property~\ref{prop:disconnect} is satisfied because a node can only be disconnected if it is marked. 
If $S$ is a set of marked nodes with an unmarked parent $P$,
$S$ can be disconnected by a CAS that swings $P.next$ from pointing to the first node of $S$ to pointing to the node after the last node of $S$.
This is the only legal instruction on $P$ that is able to disconnect exactly the nodes in $S$, so Item \ref{item:uniqueDisconnect} of Property~\ref{prop:disconnect} is satisfied.
If several nodes are marked, removing them in any order yields the same result: a list with all of the marked nodes removed, and all of the rest of the nodes still connected in the same order.
Thus, all items of Property~\ref{prop:disconnect} are satisfied.


\subsection{Algorithmic Supplements}\label{subsec:supplements}

We now present two additional requirements for \ourDS{s}. These requirements are imposed so that a \ourDS{} can go through the transformation to being persistent. The first supplement that we require is a function that disconnects all marked nodes from the data structure, and the second is an additional field that keeps information to be used by the transformation. We do not expect these properties to naturally appear in a lock-free algorithm and we therefore call them `supplements'. They should be added to a data structure for it to become a \ourDS. Both supplements are easy to implement. 
 

\begin{supplement}\label{supp:disconnect}
	\he{There is a function $disconnect(root)$ which takes in the root of the \ourDS{} and satisfies the following properties:}
	\begin{enumerate}
		\item \he{$disconnect(root)$ can be run at any time during an execution of the \ourDS{} (without affecting the linearizability of the \ourDS{}).}
		\item \he{$disconnect(root)$ can only perform disconnect instructions defined in Item \ref{item:uniqueDisconnect} of Property \ref{prop:disconnect} and no other modifying instructions.}
		\item \he{If no \ourDS{} operation takes a step during an execution of $disconnect(root)$, then there will be no marked nodes at the end of the $disconnect(root)$.}
	\end{enumerate}
\end{supplement}

The $disconnect(root)$ operation can be implemented by traversing the data structure and using the the unique atomic disconnection instruction for the marked nodes. 
For our running example of Harris's linked list, we can supply a function that traverses the linked list from the root pointer and trims all the marked nodes. 


The second supplement that we require for a \ourDS{} is that it keeps an extra field in each node, which stores the {\em original parent} of this node in the data structure. Since the data structure is a tree, a node can only have a single parent when it joins the data structure. We require the address of the pointer field that was changed to link in the new node to be recorded in the extra field. 
Note that it is possible that a sub-tree is added as a whole by linking it to a single (parent) pointer in the data structure. In this case, that same parent pointer should be stored in all the nodes of the inserted sub-tree.  The location of this pointer must be stored in the original parent field \emph{before} the node is linked to the data structure to ensure that this field is always populated. 

\begin{supplement}\label{supp:origparent}
	A designated field in each node $n$, called \emph{the original parent} of $n$, must store the location of the pointer that was used to connect $n$ to the data structure. 
\end{supplement}

In ~\Cref{sec:flushes} we specify how this field is used. Adding a field to the data structure may be space consuming, so we also propose an optimization that can avoid storing this field. 

In our running example, before inserting a node to Harris's linked list we put the address of the next field of the preceding node in the \emph{original parent field} of the new node.

	\renewcommand{\figurename}{Algorithm}
\begin{figure}
\caption{Operation in an \durableOurDS}
\begin{lstlisting}[frame=single, language=c++]
@$\textbf{T}$@  operation (@$\textbf{Node}$@ root, @$\textbf{T'}$@  input) {
	while (true) {
		@$\textbf{Node}$@ entry = @\textbf{\findentry}@ (root, input);
		List<@$\textbf{Node}$@> nodes = @\textbf{\traverse}@ (entry, input);
		@\textbf{\reach}@ (nodes.first());
		@\textbf{\makePersist}@ (nodes);
		bool restart, @$\textbf{T}$@ val = @\textbf{\critical}@ (nodes, input);
		if (!restart) 
			return val;	
	}	
} \end{lstlisting}
	\label{alg:durableOurDS}
\end{figure}

\section{\DurableOurDS}\label{sec:flushes}

In this section we show how to apply flush and fence instructions to any \ourDS{} to create an efficient and provably correct durably linearizable version of it. These flush and fence instructions can be applied \emph{automatically}

At a high-level, no persisting is done during the \traverse{} method, whereas, in the \critical{} method, every field accessed must be persisted before the next \modret{} instruction is executed. Furthermore, we add another phase between the \traverse{} and \critical{} methods, in which we ensure that the nodes returned by the \traverse{} method are persisted.

\textbf{Recovery.} \Michal{
The recovery phase executes the disconnection function guaranteed by \Cref{supp:disconnect} in \Cref{subsec:supplements}. No additional action is required.
}

\subsection{Before the Critical Method}\label{subsec:beforecritical}

We now specify the fields that must be persisted before the critical method begins. 
\begin{protocol}\label{prot:trav}
	Let $n_1 \ldots n_k$ be the nodes that were returned by the \traverse{} method of some operation $op$, where $n_1$ is the topmost node returned.
	Before the beginning of the \critical{} method of $op$, the following fields must be persisted.
	\begin{itemize}
		\item The original parent pointer of $n_1$.
		\item All fields that the \traverse{} method read in $n_1 \ldots n_k$.
	\end{itemize}
\end{protocol}

We flush these fields in two functions, called \emph{\reach} and \emph{\makePersist}, corresponding to the first and second items, respectively.
We briefly describe how we implement each of these functions. Note that these functions are the same for all \ourDS{s}, and can simply be inserted as black boxes between the \traverse{} and \critical{} methods of a given \ourDS.

\textbf{\reach.} The \reach{} function's goal is simply to flush one field: the original parent pointer of the topmost node returned by the \traverse{} method. Note that the \emph{original} parent of a node might not be the \emph{current} parent of that node, since other nodes may have been inserted in between. 
By \Cref{supp:origparent} (from \Cref{subsec:supplements}) the original parent field is available in the node. EnsureReachable gets the first node from the traversal as an input and flushes the location recorded in its original parent field.

\textbf{An optimization for \reach.} While the proposed original parent mechanism is simple, it can also be costly, since it requires an extra word on each node, and may also delay garbage collection. 
We therefore present an alternative solution for the common case that the insert operation of the data structure always connects a single node to the structure. In this case, \reach{} may simply flush the \emph{current} parent of its input node. To implement this alternative, the traversal phase returns one extra node, which is the \emph{current} parent of the first node  returned from the traversal. 

This method can also be used if the insert operation links at most $k>1$ nodes to the structure simultaneously, but becomes less efficient. In this case, the traversal needs to return the last $k$ nodes on the traversal path towards the first node that the traversal procedure returns. These nodes are then flushed by \reach{}. 
In ~\Cref{subsec:correctnessNVTraverse}, we prove this approach correct. 

We summarize this in the following lemma.

\begin{lemma}\label{lem:parent}
	In an \durableOurDS{} implementation in which the deepest tree ever atomically inserted is of depth $k$, the \reach{($n$)} method can be implemented as follows.
	\begin{itemize}
		\item \label{item:reachOriginal} If $n$ has an OP field, flush the location in this field.
		\item Otherwise, flush a path of length $k$ back from $n$.
	\end{itemize}
\end{lemma}

To prove this lemma, we start by proving the following:

\begin{lemma}\label{lem:deleteFlush}
	Let $n$ be a node and $n_d$ be some descendant of $n$. Furthermore, let $c_{n,d}$ be the child pointer of $n$ the points to the subtree in which $n_d$ is. If $n$ is deleted from the data structure at time $t$ and $n_d$ remains in the data structure at time $t$, then $c_{n,d}$ is flushed before time $t$.
\end{lemma}

\begin{proof}
	We first show that there must have been some high-level operation that read an ancestor of  $n_d$ at or below $n$, and made a change at or above $n$. 
	Let $n$ be a node that is deleted from the data structure at time $t$ by the high-level operation $op_{del}$.
	Consider child $c_{n,d}$ of $n$, and let $T$ be the subtree rooted at $c_{n,d}$, such that $T$ is not completely deleted at time $t$. Let $n'$ be the topmost non-deleted node in $T$ at time $t$. Note that for $n'$ to remain undeleted in the data structure while $n$ is deleted, the deletion \emph{must} occur by swinging a pointer from some ancestor of $n$ to some (new) node that points to an ancestor of $n'$ (or $n'$ itself). This is because of the tree property (Property~\ref{prop:tree}); $n'$ necessarily has only one parent, and for it to remain connected while its parent is removed, a node that is connected to the data structure must point to $n'$ immediately after the deletion occurs. Therefore,  $op_{del}$ mush have known $n'$'s or $n'$'s ancestor's address. However, this info is stored at or above $n'$, and in particular at or below $n$. According to Property~\ref{prop:procIndependent}, the only information an operation has is its parameters, which contain only one entry point to the data structure. Therefore, in order to know $n'$'s address (or its ancestor's address), either $op_{del}$ read it from below $n$ itself, or some other operation wrote it above $n$. Consider $op$ was the first operation that wrote $n'$'s address somewhere above $n$. Then $op$ must have read it below $n$, proving our claim. 
	
	Therefore, $op$ must have read the child pointer of $n$ that points to $n'$'s subtree, because this is the only way it could have read $n'$'s (or its ancestor) address and store it above $n$. Recall that by the way the \OurDS{s} operations work, for $op$ or $op_{del}$ to read something below $n$ and subsequently write above $n$, there are only two possibilities: 
	\begin{enumerate}
		\item The read below $n$, including $c_{n,d}$ was during $op$'s or $op_{del}$'s \traverse{} method, but below the its return point.
		After reading $c_{n,d}$, those operations went on to make a change at or above $n$. Thus, $op$'s or $op_{del}$'s \traverse{} method must have returned a node at or above $n$.
		\item The read below $n$, including $c_{n,d}$ was during the \critical{} method.
	\end{enumerate}	
	
	Thus, in both these cases, $c_{n,d}$ must have been flushed; either in the \makePersist{} method, or during the \critical{} method.
\end{proof}

We now continue with proving Lemma~\ref{lem:parent}.

\begin{figure}[h]
	\centering
	\caption{Diagram for Case 1 of \Cref{lem:parent}}\label{fig:ensureReachCase1}
	\includegraphics[width=0.8\columnwidth]{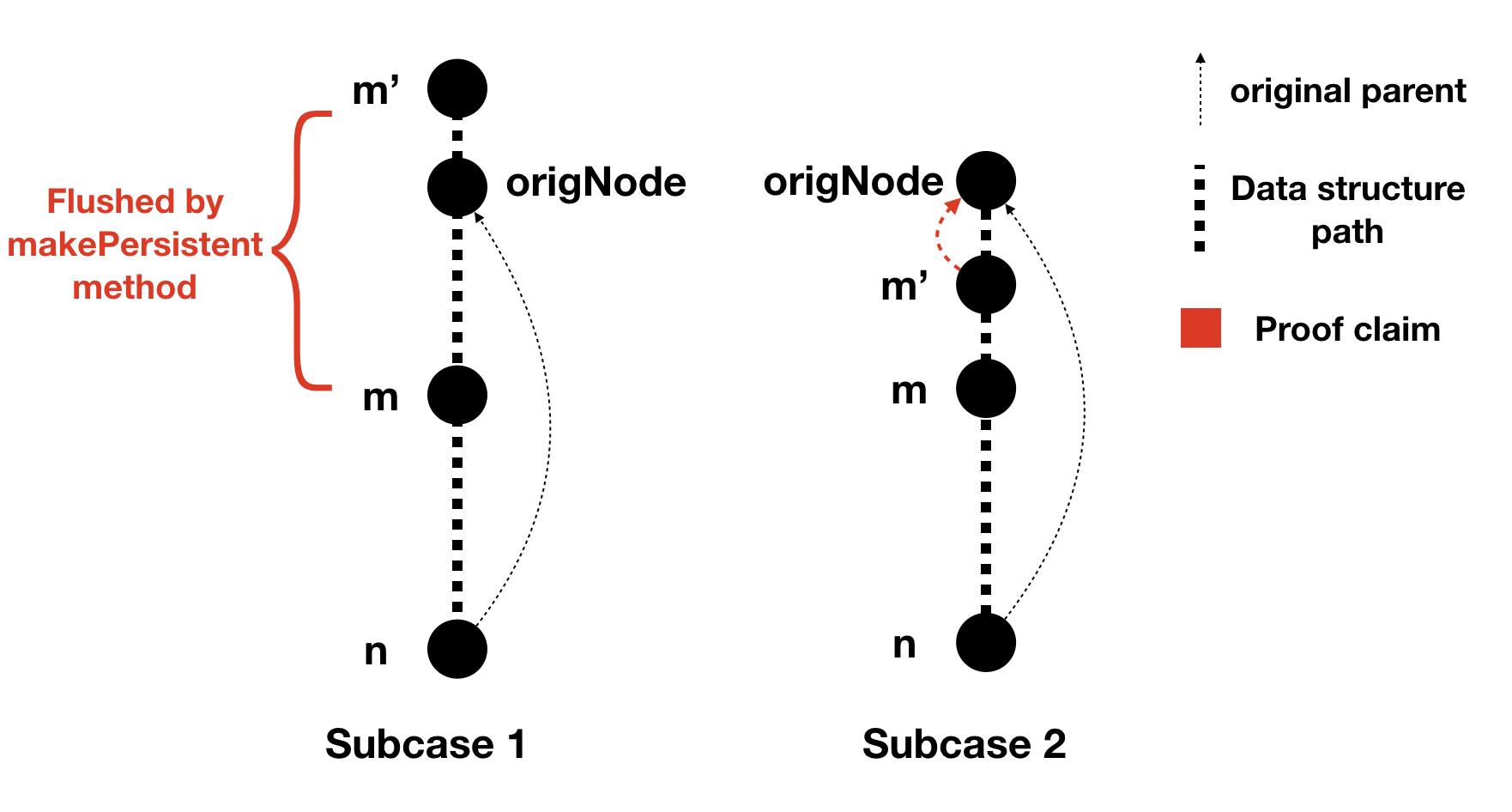}
\end{figure}

\begin{proof}
	Assume by contradiction that the lemma does not hold. That is, there is some node $n$ returned by some \traverse{} method such that, after the execution of \reach{} with the implementation specified by the lemma, $n$'s original parent pointer has not been flushed.
	Let $n$'s original parent \emph{node} be $origNode$, and let $origPtr$ be the child field of $origNode$ on which the instruction that atomically inserted $n$ into the data structure occurred.

	If $n$ had an originalParent field, then by the first option of the \reach{} method implementation, $n$'s originalParent field would have pointed to $origPtr$, and $origPtr$ would have been flushed. Contradiction.
	
	Otherwise, if $n$ does not have an originalParent field, the \reach{} method would have flushed a path of length $k$ up from $n$. Since we assumed that $n$'s original parent pointer, i.e., $origPtr$, was not flushed by the end of the \reach{} method, this must mean that $origNode$ was not in one of these $k$ nodes. However, recall that by the definition of $k$, $n$ could not have been inserted at a distance more than $k$ away from its original parent node. Therefore, at least one of the following must have happened: (1) there was at least one insert operation at some node between $n$ and $origNode$ since $n$ was inserted, or (2) $origNode$ was deleted from the data structure before this \reach{} method was executed. We handle these two cases separately. 
	
	\textsc{Case 1.} There was an insert operation somewhere between $n$ and $origNode$, but $origNode$ was not removed from the data structure. Let $m$ be the node at which the first such insertion occurred. The insertion at $m$ must have been a part of some high-level operation $op$, whose \traverse{} method returned either $m$ or a node above $m$. Let $m'$ be the node returned by the \traverse{} method of $op$. Consider two subcases, gudied by \Cref{fig:ensureReachCase1}:
	
	\textsc{Subcase 1.1.} If $m'$ is above $origNode$, then $op$ must have traversed the only path between $m'$ and $m$ and flushed all nodes in between (either inside its \traverse{} method or inside its \critical{} method). Thus, since by assumption $origNode$ was not deleted from the data structure before this occurred and was on the path between $m'$ and $m$ (Property~\ref{prop:tree}), $origNode$ would have been flushed by $op$ before the insertion at $m$ occurred, and therefore also before the end of the \reach{} method in question. Contradiction.
	
	\textsc{Subcase 1.2.}  If $m'$ is below $origNode$ (and above $n$), then note that since $m$ is the first node on which the insertion between $n$ and its original parent occurs, $m'$ must have been inserted in the same batch as $n$, and thus $m'$'s original parent must also be $origNode$, and its original parent pointer also $origPtr$. Note that \reach{} must be run starting on $m'$ before the insertion at $m$ can occur. Note also that by the definition of $k$ as the maximum batch depth, $m'$ was inserted at a distance smaller than $k$ from $origNode$, and at the time that \reach{} is run on $m'$, it is still at the same distance from $origNode$. This is because of the assumption that $op$ is the first insertion between $n$ and $origNode$. Thus, regardless of whether $m'$ has an originalParent field or not, the \reach{} method of $op$ flushed $origPtr$. Contradiction.
	
	\textsc{Case 2.} $origNode$ was deleted from the data structure. Then by \Cref{lem:deleteFlush}, since $n$ remains in the data structure after the deletion of $origNode$, $origPtr$ must have been flushed before $origNode$ was deleted. 
\end{proof}

\textbf{\makePersist.} The purpose of \makePersist{} is to ensure that all fields read by the traversal on the nodes that it returned are persisted. This can be done by simply flushing all of these fields, and finally executing a single fence instruction. Note that this fence also ensures the completion of the flush of  \reach{}.

\subsection{During the Critical Method} Simply put, all fields accessed by some process $p$ during the \critical{} method must be persisted before the next \modret{} instruction is executed by $p$. However, this can be relaxed for fields that cannot be accessed by any process other than $p$. Intuitively, the idea is that $p$ must ensure, before executing an instruction $e$ that may affect other processes, that the values that $p$ relied on to determine $e$'s parameters must be persisted. To achieve this, we use the following rules.

\begin{protocol}\label{prot:crit}
	In any \critical{} method, the following flush and fence instructions must be injected.
	\begin{itemize}
		\item Flush after every read of a shared variable.
		\item Flush after every write/CAS instruction.
		\item Fence before every write/CAS on a shared variable.
		\item Fence before every return statements.
	\end{itemize}
\end{protocol}

Note that this means that local work requires less flush and fence instructions than shared work. In particular, when initializing a node, a process executes flushes after initializing each field, but only needs to fence once before atomically inserting the new node into the data structure. Furthermore, there is no need to flush after reading an immutable field.

\subsection{Correctness}\label{subsec:correctnessNVTraverse}

We say that the algorithm resulting from applying \Cref{prot:trav}, \Cref{prot:crit} and the specified recovery procedure to a \ourDS{} form an \emph{\durableOurDS}. The algorithm is  presented in Algorithm~\ref{alg:durableOurDS}. We continue with the proof of the following theorem:

\begin{theorem} \label{thm:durable}
	Every \durableOurDS{} is durably linearizable.
\end{theorem}

Intuitively, a data structure is durable linearizability if a crash cannot make the effect of any operation be lost. \durableOurDS{s} achieve this by ensuring that every modification to shared memory is persisted before any process uses that modification's value. That is, every value read by $p$ is persisted before $p$ makes its next shared memory modification. The only exception is during the traversal, where a process may read values that it does not know are flushed. However, due to the restrictions on how the \traverse{} method may behave, no modification that $p$ does, can depend on a value it read (but did not return) during its traversal. Other than persisting all values that may affect $p$'s shared memory modifications, we also need to make sure that $p$'s writes cannot disappear from the data structure at a later point. This can happen if $p$ wrote on a part of the data structure that might not be \emph{reachable} from the root
upon a crash. To prevent this from happening, the \reach{} function persists the pointer that connects the relevant subtree to the rest of the structure. Thus, the flush and fence instructions we prescribe are \emph{necessary}; removing any of them could violate the correctness of some \durableOurDS. However, hand-tuned data structure implementations could still save on flushes and fences by reasoning more carefully about dependencies in the data.

We now show formally that a \durableOurDS{} is durably linearizable~\citep{izraelevitz2016linearizability}.
That is, we show that the flushes described in Section~\ref{sec:flushes} are sufficient to ensure that any high-level history $H$ of a \durableOurDS{} with any number of system crashes is linearizable if we remove the crash and recovery events from it.

Consider a legal low-level execution $H$ of a \durableOurDS.
In the proof, we say that a successful modifying instruction $m$ \emph{was seen} by some instruction $i$ if $i$ was executed on the same location $\ell$, and was after $m$ in $H$.
Furthermore, we say that an instruction $i$ is \emph{below (resp. above)} another instruction $i'$ if $i$ is executed on a node $n$ such that the path from the root of the data structure to $n$ passes through the node $n'$ on which $i'$ was executed (resp. $n'$ is on the path from the root to $n$).



To help with our proof of durability, we first prove that the recovery function can be quite useful to us. More specifically, we show that the recovery function will always execute any pending \disconnect{} instruction. For this purpose, we first introduce the notion of \emph{dependence} between low-level instructions that modify data, and relate that notion to flush ordering.

\begin{definition}
	Let $H$ be a legal execution of a \durableOurDS{}, and $m$ be a modifying instruction in $H$ that was executed in the high-level operation $op_m$.
	We say that $m$ \emph{depends on} another modifying instruction $m'$ in $H$ if one of the following conditions hold.
	\begin{enumerate}
		\item  There was a read instruction $r$ in $op_m$ such that $r$ saw $m'$, and the \traverse{} method of $op_m$ returned a node at or above $r$.
		\item $m$ and $m'$ were executed by the same process, and $m'$ is before $m$ in $H$.
	\end{enumerate}
\end{definition}

Here the idea is that if a modifying instruction $m$ depends on another modifying instruction $m'$, then we can show that $m'$ must have been flushed before $m$ was executed. Furthermore, we will show that the extended notion of dependency (formally defined below as a \emph{dependency graph}) captures all instructions that might have affected the decision of $m$'s parameters, that is, where and when it is executed, as well as what exactly is written. These two insights together intuitively imply that if we were to lose the effect of a modifying instruction that was not yet flushed, we would be able to recreate that instruction by observing the state of the data structure.

\begin{definition}
	The \emph{dependency graph} of an execution $H$ of a \ourDS{} is a directed graph in which modifying low-level instructions are vertices, and there is a directed edge from $m$ to $m'$ if $m$ depends on $m'$ in $H$.
	For a modifying operation $m$ in $H$, the \emph{dependency graph of $m$} is the subgraph reachable from $m$ in the dependency graph of $H$.
\end{definition}

\begin{claim}\label{clm:dependencyFlush}
	Let $m$ be a modifying instruction in a history $H$ of a \ourDS{}. Then all instructions in $m$'s dependency graph, other than $m$ itself, were flushed in $H$ before $m$ was executed.
\end{claim}

\begin{proof}
	Consider a BFS tree $T$ of $m$'s dependency graph. Note that $m$ is the root of $T$.
	The proof is by induction on the number of levels in $T$. Let $op_m$ be the high-level operation that executed $m$.
	
	\textsc{Base:} If $T$ is a singleton, then there are no modifying instructions in $T$ other than $m$ itself, so the claim is vacuously true. 
	
	\textsc{Step:} Assume that for any modifying instruction $m'$ for which the depth of the BFS tree of its dependency graph is at most $k-1$, all instructions in $m'$'s dependency graph were flushed before $m'$ was executed.
	Then consider a modifying instruction $m$ whose dependency graph is of depth $k$. 
	
	Let $T$ be a BFS tree of $m$'s dependency graph. The depth of the dependency graphs of the children of $m$ in the tree is at most $k-1$. Let $c$ be a child of $m$ in $T$, and let the subtree of $T$ rooted at $c$ be $T_c$. By the induction hypothesis, all nodes in $T_c$ are flushed before $c$ is executed, other than $c$ itself. Therefore we only need to consider, for each child $c$ of $m$, whether $c$ is flushed before $m$ was executed in $H$. So, first note that $m$ is executed after $c$ is executed. Furthermore, note that by the definition of $T$, $m$ depends on $c$. By the definition of dependence, $c$ was either (1) seen by a read instruction in $op_m$ that was at or below the return point of the \traverse{} method of $op_m$, or (2) executed by the same process as $m$, before $m$. 
	In both these cases, $c$ must have been flushed before executing $m$; in the first case, $op_m$ would have flushed $c$ either during the \makePersist{} method, or in the critical section, and in the second case, $p$ must have flushed $c$ before proceeding to its next instruction in the critical section.
\end{proof}

To show that the recovery can execute all pending deletion instructions, our goal is to show that if a deletion instruction is pending, then the state of the data structure after a crash would identify this deletion instruction. Recall that by Property~\ref{prop:disconnect}, for every given set of contiguous marked nodes $S$, and an unmarked $Parent(S)$, there is exactly one legal disconnection instruction that can be executed on $Parent(S)$.
Furthermore, once $S$ enters such a state, no node in $S$ can change.
We therefore begin by showing that before a deletion instruction can be executed, all modifications of nodes that mark $S$ must be flushed.

\begin{claim}\label{clm:flushM}
	Let $d$ be a deletion instruction and let $S$ be the set of nodes that it deletes. Furthermore, let $H$ be a legal history in which $S$ enters a state in which it is marked, and let $M$ be the set of modifications on nodes that mark these nodes. Then before $d$ is executed in $H$, all instructions in $M$ have been flushed.
\end{claim}

\begin{proof}
	Assume by contradiction that the claim does not hold. Then there is at least one modifying instruction, $m\in M$, without which $S$ would not be marked, but which was not flushed before the execution of $d$ in $H$. Let $n$ be the node that $m$ operated on. Let $p$ be the process that executes $d$ in $H$.
	Then by Claim~\ref{clm:dependencyFlush}, $m$ cannot be in the dependency graph of $d$. Moreover, $m$ cannot be in the dependency graph of \emph{any} modification instruction in $H$, since otherwise it would have been flushed. 
	We now construct a history $H'$ which is legal, and has the following two properties: (1) there is no instruction that executes $m$ on $n$, and (2) $d$ takes place.

	Consider all processes $q_1 \ldots q_i$ that attempt to execute $m$ on $n$ in $H$. Without loss of generality, assume that $q_1$ executes $m$ in $H$, and that all other $q_j$'s execute some failed CAS $c_j$.
	Then in $H'$, we pause all these processes before they execute this modification attempt.
	Furthermore, we treat any read instruction $r$ that read $m$ in $H$ as follows:
	if the operation in which $r$ was executed did not execute any modification instructions or return statements after $r$, then we pause $op$ before executing $r$. Otherwise, we reschedule any $r$ such that it returns the previous value at that location.
	Clearly, $H'$ satisfies properties (1) and (2) specified above, since we remove all attempts to execute $m$ on $n$ in $H$, and do not remove any other modification (in particular, we do not remove $d$).
	We now show that $H'$ is legal.
	
	Since by assumption, $m$ was not flushed in $H$, no process in $q_1 \ldots q_i$ could have executed any modification instruction after their attempt to modify $n$. This is because any such attempt would have been flushed before the next modification or return statement by that process, thereby flushing $m$.
	Thus, pausing these processes before their attempted modification of $n$ cannot affect any other process in the execution, except possibly the processes that saw $m$.
	We therefore only have to argue that replacing a read that saw $m$ with a read that saw the previous value at that location does not affect any other instruction in the execution.
	
	Let $r$ be a read instruction that saw $m$, and let $op$ be the operation that executed $r$.
	Recall that $r$ must have either (1) been during the \traverse{} method of $op$, and the \traverse{} returned somewhere \emph{below} $m$, or (2) not been followed by any modification or return statement in $op$. This is because otherwise, $m$ would have been flushed by $op$.
	Case (2) is easily treated since by the definition of $H'$, $op$ is paused before executing $r$, and since it did not modify anything else, it cannot affect any other process.
	So consider a read instruction $r$ that saw $m$ in the \traverse{} method of $op$, and the \traverse{} returned below $m$. By \Cref{item:stopTraversal} and \Cref{item:returnTraversal} of Property~\ref{prop:traverse}, $m$ could not have made $op$'s traversal stop, since then the traversal would have returned at or above $m$. By \Cref{item:directionTraversal} of Property~\ref{prop:traverse}, $m$ could not have affected the direction in which the traversal proceeded, since that decision is made only using immutable fields. Finally, recall that since $m$ was executed on $n$, $m$ must not have been an insertion or a deletion or a node, and must only have modified the values of $n$. By \Cref{item:stableTraversal} of Property~\ref{prop:traverse}, if the \traverse{} returned below $m$, a \traverse{} that had not seen $m$ could not have stopped at $n$, meaning that $m$ could not have been the reason for $op$ to continue its traversal. Therefore, reading the previous value instead of $m$ has no effect on the traversal of $op$, and therefore no effect on the rest of $op$ either.
	
\end{proof}

We are now ready to show that the recovery function can execute \disconnect{} instruction if we lose them upon a crash.

\begin{lemma}\label{lem:recover}
	After a crash event, the recovery function of a \durableOurDS{} executes all disconnect instructions that were pending at the time of the crash.
\end{lemma}

\begin{proof}
	Let $H$ be a legal history of a \durableOurDS{} algorithm $A$, in which $m$ is a modifying instruction that \disconnect{ed} the set of nodes $S$ from the data structure.
	Let $p$ be the process that executed $m$. 
	By \Cref{item:uniqueDisconnect} of Property~\ref{prop:disconnect}, there is a set of contiguous nodes $S$, such that $m$ is legal on $Parent(S)$ only if $S$ was marked.
	So, consider the set of modifying instructions, $M$, that marked $S$.
	By Claim~\ref{clm:flushM}, all instructions in $M$ were flushed in $H$ before $m$ was executed. 
	
	Note that by 
	Property~\ref{prop:disconnect}, once $S$ is marked, all nodes in $S$ stay in their state. Furthermore, note that by Property~\ref{prop:tree}, each node in a \ourDS{} only has one incoming edge. Therefore, there is exactly one pointer that must be swung in order to atomically disconnect $S$; this is a pointer that resides in node $Parent(S)$, and thus $m$ must be a modification on this pointer. By Claim~\ref{clm:flushM}, at the time that $m$ occurs in $H$, all the marks has been flushed. Note also that since $m$ was still pending at the time of the crash, all the nodes must still have been marked since no node in $S$ could have changed (by  Property~\ref{prop:disconnect}). Moreover, if any external mark was reverted, by Item~\ref{item:removeMark} of Property~\ref{prop:disconnect}, it must have been done after executing $m$, failing to execute $m$ because another thread has done it earlier, or reading the location on which $m$ was executed to disconnect $S$. All these operations must have been made in the critical section, since updates are done only in the critical section. In addition, reading the location on which $m$ was executed to disconnect $S$, which leads to reverting the external marks must also been done in the critical section due to Property~\ref{prop:traverse}. Therefore, in this case, it means that $m$ was flushed too, contradicting the assumption that $m$ was pending.
	Therefore, the recovery function finds $S$ marked, and thus by Supplement~\ref{supp:disconnect}, it executes $m$.
\end{proof}

\begin{lemma}\label{lem:main}
	Let $H_r$ be a low-level history of a \durableOurDS{} with exactly one crash event at the end of the history, followed by a recovery. 
	There exists a low-level history $H_r'$ (without crashes or flushes) such that:
	\begin{enumerate}
		\item $H_r'$ is a legal execution of the same data structure
		\item $H_r$ is equivalent to $H_r'$ 
		 \item $H_r'$ contains exactly all the flushed modifying instructions in $H_r$ in the same order. $H_r'$ does not contain any pending modifications. \label{prop:writes}
	\end{enumerate}
\end{lemma}


\begin{proof}
	
	We prove the lemma constructively; given history $H$, we show how to construct a history $H'$ that satisfies the necessary properties.
	We do so iteratively by considering one pending modification at a time, starting with the last such modification in $H$ and working backwards.
	
	Let $w_1$ be the last pending modifying instruction in $H$, and let $p_1$ be the process that executed it, and $\ell$ be the location on which it was executed.
	First note that by \Cref{lem:recover}, if 
	$w_1$ disconnected a set of nodes, $S$, from the data structure, then the recovery function can redo $w_1$ at any order by \Cref{item:overlapDisconnect} after a crash, thus $w_1$ is not pending.
	We therefore leave $w_1$ in $H'$ in this case. 
	
	We now consider several cases, depending on what operations follow $w_1$. Intuitively, in each case we either show that we can remove $w_1$ from the history without harming its legality, or that $w_1$ must have already been flushed, contradicting the assumption that it is pending.

	\smallskip\noindent
	\textsc{Case 1.} Suppose $w_1$ was seen by no low-level instruction. Then we can remove $w_1$ from $H$.
	
	\smallskip\noindent
	\textsc{Case 2.} Suppose $w_1$ was seen by some modification $m_2$ by process $p$.  If $m_2$ was flushed (i.e., it is not pending), then $w_1$ was flushed as well, which is a contradiction. 
	If $m_2$ was not flushed then $w_1$ was not the last pending modification, which contradicts the choice of $w_1$.
	
	\smallskip\noindent
	\textsc{Case 3.} Suppose $w_1$ was seen by some read instruction. Let $P$ be the set of processes that had read instructions that saw $w_1$ in $H$. For each such process $p$, let $r$ be the read that saw $w_1$. Consider the following subcases.
	\begin{enumerate}
		\item There is no modifying instruction or return statement by $p$ after $r$ in the same high-level operation attempt in $H$. Then we can remove $r$ and all subsequent instructions of the same operation attempt by $p$. If $p$ has subsequent operations or repetitions of this operation, in $H'$ we let a new process, $p'$ execute these. This leads to a legal execution history since no other process could have been affected in $H$ by $p$'s reads, and since no information is passed between operation attempts, and thus the new process $p'$ can behave exactly as $p$ would have as $p'$ can get the same arguments as $p$ in its following attempt. This is equivalent to a history in which $p$ stalls right before executing $r$. 
		\item Suppose there is some modification $m$ or return statement $R$ by process $p$ after $r$, but $r$ was either (1) inside a \critical{} method, or (2) inside a \traverse{} method, but below the node that this \traverse{} method returned. In both cases, before $p$ executes $m$ or $R$, it must flush the location at which $r$ executed, and therefore, $p$ flushes $w_1$ as well. This contradicts the assumption that $w_1$ is pending.
		
		\item Suppose there is a modification, $m$, or return statement, $R$, by process $p$ after $r$ in the same operation attempt as $r$, and $r$ was in a \traverse{} method that returned a node below $r$.
		%
		%
		Note that the modification/return statement must have happened inside the \critical{} method, since by the definition of \ourDS, all modifying instructions and return statements are in the \critical{} method. 
		Recall that $w_1$ could not have disconnected any nodes, since we've already covered this case. We thus consider two cases: 
		(1) $w_1$ inserted some set of new nodes $S$, but did not \disconnect{} any nodes, or (2) it neither inserted nor \disconnect{ed} any nodes.
		
		
		\smallskip
		\textsc{Case 1.} $w_1$ inserted a set $S$ of new nodes into the data structure. 
		Then consider two subcases, depending on the location of the node $n$ returned by the \traverse{} method of $p$ in which $r$ was executed: 
		
		\textsc{Subcase 1.} $n\in S$. That is, $p$'s \traverse{} method in which it read $r$ returned one of the new nodes that $w_1$ inserted. Then recall that between every pair of \traverse{} and \critical{} methods in the same operation, we run a \reach{} function that ensures that the original parent of the node returned by the \traverse{} is flushed. Note that in this case, $n$'s orignal parent pointer location is $\ell$ (where $w_1$ occurred). Therefore, before the \critical{} method of $p$'s operation begins, $\ell$ must have been flushed, and thus $w_1$ is no longer pending. Contradiction.
		%
		
		\textsc{Subcase 2.} $n \not\in S$. That is, the node returned by $p$'s \traverse{} method is not part of the set of new nodes added by $w_1$. 
		
		Let $n_1$ be the node that $w_1$ operates on and $n_2$ be the node pointed to by the previous value of $w_1$. We first claim that the traversal in $H$ must have passed through $n_2$. We know that the traversal passes through $n_1$ and it does not stop on any node in $S$. Therefore it must eventually reach a node that is not in $S$. Let $n_3$ be the first such node. If $n_3$ was not in the data structure immediately before $w_1$ then it must have been inserted by a pointer swing on some node in the set $S$. However due to the flushes we add, all the nodes in $S$ have to be reachable by traversing persistent memory before the pointer swing can occur. This would contradict the fact that $w_1$ is pending. Therefore $n_3$ must have been in the data structure immediately before $w_1$. Now suppose for contradiction that $n_3 \neq n_2$. Then that means there are two paths from the root to $n_3$, which violates Property~\ref{prop:tree}. This is because there was a path to $n_3$ before $w_1$, and $w_1$ adds a new path to $n_3$. Therefore $n_3 = n_2$ which means the traversal passes through $n_2$. So we can pause $p$ in $H'$ immediately before its traversal visits $n_2$ and we can unpause $p$ when $p$ is about to visit $n_2$ in $H$. Since $p$ ends up at the same node in both histories and the decision of whether or not to stop and which node to visit next (Property~\ref{prop:traverse}) depends only on information inside the current node, after visiting $n_2$, $p$ would perform the exact same steps in both $H$ and $H'$. This means $p$'s interpreted history remains the same in $H$ and $H'$ and since we only removed some shared memory reads from $H$, the order of writes in $H$ are maintained.

		\smallskip
		\textsc{Case 2.} $w_1$ neither inserted nor deleted any node. Recall that the \traverse{} method's direction decision may not depend on mutable values, and thus, the \traverse{} of $p$ may not depend on the value read by $r$. Furthermore, by \Cref{item:stableTraversal} of Property~\ref{prop:traverse}, since we know that the \traverse{} method returns a node below $w_1$ in $H$, the \traverse{} method in $H'$ must stop at a node below $w_1$, meaning that reading the previous value instead of $w_1$ has no affect on the traversal in $H'$. Thus, we can safely remove $w_1$ from the history, and replace $r$ by a read instruction that returns the previous value of $\ell$.
	\end{enumerate}
	
	Therefore, for every $p$ that executed a read operation $r$ that saw $w_1$, we can safely remove $w_1$, and possibly have to change some of $p$'s subsequent reads, but not any of its subsequent modifications.
	
	To complete the proof, we need to show that the $H'$ that we've constructed satisfies the 3 properties required by the lemma.
	Firstly, note that, as agrued above, all changes and removals done to $H$ in the construction of $H'$ leave a legal history. In a nutshell, this is due to the fact that no low-level instructions from the \critical{} method are changed at all between $H$ and $H'$ other than pending modifications, which are never followed by any other step by the same process. Thus, decisions like whether or not to restart an operation are unaffected.
	Secondly, note that we never remove or change a return statement, and therefore, we do not remove or change any responses of high-level operations. Thus, the interpreted histories of $H$ and $H'$ on completed operations are the same. Consider the interpreted history of $H'$ in which, after every high-level response which is not already followed by a high level invocation, we place the invocation of the high-level operation that is invoked in $H$ at this location. We arrive at the same interpreted history for both $H$ and $H'$.
	Finally, note also that we never change or remove any completed modifying instruction in the above construction of $H'$.
	
\end{proof}

To complete the proof of \Cref{thm:durable}, let $H$ be a low-level history of a \durableOurDS{}, with one crash event followed by a recovery at the very end of the history.
Recall that, after a crash, the state of volatile memory is lost, and the state of persistent memory remains. That is, each memory location's value becomes the value of the most recent flushed write on this location. Note that by \Cref{lem:main}, there exists a legal low-level history $H'$ with no crashes such that the state of memory after $H'$ is the same as the state of memory after $H$'s crash and recovery. Furthermore, $H$ and $H'$ are \emph{equivalent}; their interpreted histories are the same. Thus, future operations cannot `see' the difference between the two histories. We formalize this intuition to complete the proof.

\begin{proof}[Proof of Theorem~\ref{thm:durable}]
	Let $H = H_1 \cdot C \cdot R \cdot H_2 \cdot C \cdot R \ldots \cdot H_n$ be a high-level history of a \durableOurDS{}, where each $H_i$ contains no crash events, and $m$ and $R$ correspond to a crash event and an execution of \durableOurDS{}'s recovery function, respectively. Furthermore, for any pair of subhistories $H_i$ and $H_j$, the set of processes in $H_i$ and $H_j$ is disjoint.
	To show that a \durableOurDS{} $D$ is durable linearizable, we must show that we can remove the crash and recovery events and obtain a linearizable history of $D$.
	We show this by induction on the number of crash events $n-1$ in $H$.
	
	\textsc{Base:} If there are no crash events in $H$, then $H$ is a linearizable history of $D$ by its definition.
	
	\textsc{Step:} Assume that if $H$ has $k$ crash events in it, then we can remove the crash and recovery events from $H$ and obtain a linearizable history of $D$. 
	Now let the number of crash events in $H$ be $k+1$. Consider the prefix $G$ of $H$ that ends immediately before the $k$'th crash and recovery event of $H$. By the induction hypothesis, we remove all crash and recovery events from $G$ and obtain a linearizable history of $D$ that contains no crash events. By \cref{lem:main}, there exists a low-level history $G'$ such that $G'$ is legal and equivalent to $G$, and the state of volatile memory after $G'$ is the same as the state of persistent memory after the crash and recovery immediately following $G$. Since $G$ and $G'$ are equivalent, any legal continuation of history $G$ is also a legal continuation of history $G'$, as long as the set of processes in the continuation is disjoint from that of $G$ and of $G'$. Thus, we can remove all crash and recovery events in $H$ and get a linearizable history of $D$.
\end{proof}



\subsection{Example}

In Algorithm~\ref{alg:harris} and Algorithm~\ref{alg:harrisSearch}, we present pseudocode showing Harris's linked-list (HLL) as an \durableOurDS. Note that the \traverse{} method ends in the middle of the search function, since the search also executes some physical deletions, which are part of the \critical{} method of a \ourDS. The \traverse{} method returns the set of nodes for the \reach{} (using the \reach{} optimization) and \makePersist{} functions to flush.

\Michal{In Algorithm~\ref{alg:harris}, in lines~\ref{line:startop}-\ref{line:endop}, we show how every operation is executed. Every operation starts with finding an entry to the core tree structure. In a linked-list, the entry point is the root of the list. Therefore, \findentry{} returns the root. After that, the \traverse{} function from Algorithm~\ref{alg:harrisSearch} is called. This function returns exactly three nodes. The \emph{right} node, which is the first unmarked node in the list whose key is greater than or equal to $k$, and \emph{left} node, which is the last unmarked ancestor of \emph{right}. In addition, it returns the \emph{current} parent of \emph{left}, as described in the optimization in \Cref{subsec:beforecritical}. In line~\ref{line:ensure}, the current parent node is flushed in order to make sure that the \emph{left} node is reachable from the head, followed by line~\ref{line:persistent}, where \makePersist{} flushes the \emph{left} and \emph{right} nodes. After that, the critical part is executed; depending on the operation, we go to either \emph{insertCritical} in line~\ref{line:startinsert}, \emph{deleteCritical} in line~\ref{line:startdelete} or \emph{findCritical} in line~\ref{line:startfind} in Algorithm~\ref{alg:harrisSearch}. If those operations need to restart, they return true for the restart variable, and the operation is re-executed. 

The critical function of an insert operation, in lines~\ref{line:startinsert}-~\ref{line:endinsert} in Algorithm~\ref{alg:harris} starts by deleting marked nodes. This deletion occurs only if the \emph{left} and \emph{right} nodes that were returned from \traverse{} are not adjacent. If \emph{left} and \emph{right} were not adjacent and the deletion from lines~\ref{line:startdeletemarked}-~\ref{line:enddeletemarked} in Algorithm~\ref{alg:harrisSearch} fails, the insert operation restarts. If the key already exists, the operation returns false (lines~\ref{line:exsistinsert}-\ref{line:returnexsistinsert}). As the key is an immutable field, we do not flush after reading the key.
Note that \emph{deleteMarkedNodes} executes a fence before returning. Therefore, there is no need to re-execute that fence in lines~\ref{line:returnreinsert} and~\ref{line:returnexsistinsert}. Afterwards, a new node is allocated with the correct value, followed by a flush after write. In line~\ref{line:fencebeforeCASinsert} there is a fence before the CAS which is executed in line~\ref{line:casinsert} in order to insert the newly allocated node. This CAS is followed by a flush after CAS and a fence before the return. If the insertion has failed due to concurrent activity, the operation will be re-executed (line~\ref{line:retryinsert}). 
The critical functions of the delete and find operations follow the same rules. 

The \traverse{} function is presented in Algorithm~\ref{alg:harrisSearch} in lines~\ref{line:starttraverse}-\ref{line:endtraverse}. The inner while loop, from line~\ref{line:startinlooptraverse} to line~\ref{line:endinlooptraverse}, traverses the list from the root and tries to find the first node which is unmarked with a key equal to or greater than $k$. The marked nodes before $k$ are saved in the $nodes$ variable. After the right node is inserted to $nodes$ in line~\ref{line:apprighttraverse}, the $nodes$ variable contains the \emph{left} node which was unmarked at the moment it was inserted (in line~\ref{line:applefttraverse}), followed by all the marked nodes until the \emph{right} node (line~\ref{line:apprighttraverse}). If the right node is marked by the time line~\ref{line:rechecktraverse} is executed, the traversal restarts. If not, we proceed to line~\ref{line:parenttraverse} where we insert \emph{left's} parent to the $parent$ variable and return both the $parent$ and $nodes$ variables (to allow us to persist them later on). Note that by the given properties, no modification is ever done in the \traverse{}.

The last function we present here is called \emph{deleteMarkedNodes} (lines~\ref{line:startdeletemarked}-\ref{line:enddeletemarked} of Algorithm~\ref{alg:harrisSearch}). This function gets the nodes from the \traverse{} as an input and checks whether there are more than two nodes (more than the left and right ones). If it finds more than two nodes, then there is a need to trim all the marked ones by executing a CAS in line~\ref{line:casdeletemarked}. The key observation here that the CAS will be successful only if the current \emph{left.next} pointer is still the pointer that was read during the \traverse{}. If this is the case, the marked nodes will be trimmed successfully and the changed field will be flushed afterwards. In line~\ref{line:rightokdeletemarked} we make sure again that the \emph{right} node is not marked. If the node is marked, or the trimming was unsuccessful (line~\ref{line:enddeletemarked}), then the function will return false and the traversal will need to be re-executed. Otherwise it returns true. Before every return, we make sure that a fence is executed. Moreover, in line~\ref{line:flushdeletemarked} there is a flush which is done due to the read of the shared variable in line~\ref{line:rightokdeletemarked}.

Some further optimizations can be done, but we omit them from the pseudocode for readability.}

	\renewcommand{\figurename}{Algorithm}
\begin{figure}
	\caption{HLL Persistent Insert and Delete}
\begin{lstlisting}[frame=single, language=c++]
class Node<@$\textbf{T}$@, @$\textbf{V}$@>  {
	@$\textbf{T}$@ key; 					// immutable field
	@$\textbf{V}$@ value;
	@$\textbf{Node*}$@ next; 
}

bool operation (@$\textbf{T}$@ key) { @\label{line:startop}@
	bool restart, val = true, false;
	while (restart) {
		@$\textbf{Node*}$@ entry = @\textbf{\findentry}@(root, input);
		List<@$\textbf{Node*}$@> parent, nodes = @\textbf{\traverse}@(root, key);
		@\textbf{\emph{\flush}}@ (&parent.next); // ensureReachable() @\label{line:ensure}@
		@\textbf{\makePersist}@ (nodes); @\label{line:persistent}@
		restart, val = @\textbf{opCritical}@ (nodes, key); 
		if (!restart) 
			return val; 
	}  
} @\label{line:endop}@
	
bool, bool insertCritical(List<Node*> nodes,T key)  @\label{line:startinsert}@
	bool succDelete = @\textbf{deleteMarkedNodes}@ (nodes);
	if (succDelete == false) {
		return true, false;  // retry @\label{line:returnreinsert}@
	}
	@$\textbf{Node*}$@ left, right = nodes.front(), nodes.back();
	if (right.key == key) { // no flush - immutable @\label{line:exsistinsert}@
		return false , false; 	// key exists @\label{line:returnexsistinsert}@
	}
	@$\textbf{Node*}$@ newNode = new Node(key, right); 
	@\textbf{\emph{\flush}}@ (newNode);
	@\textbf{\emph{\fence}}@ // before CAS @\label{line:fencebeforeCASinsert}@
	bool res = CAS(&(left.next), right, newNode); @\label{line:casinsert}@
	@\textbf{\emph{\flush}}@ (&left.next); 
	@\textbf{\emph{\fence}}@; // before return 
	if (res) { 
		return false, true; //	node inserted
	} else { 
		return true, false; // retry @\label{line:retryinsert}@ 
	} 
} @\label{line:endinsert}@ 

bool, bool deleteCritical (List<@$\textbf{Node*}$@> nodes,@$\textbf{T}$@ key) @\label{line:startdelete}@
	bool succDelete = @\textbf{deleteMarkedNodes}@ (nodes);
	if (succDelete == false) { 
		return true, false; // retry
	}
	@$\textbf{Node*}$@ left, right = nodes.front(), nodes.back();
	if (right.key != key) {
		return false, false;  // no key		
	}
	@$\textbf{Node*}$@ rNext = right.next;
	@\textbf{\emph{\flush}}@ (&right.next);
	if (!isMarked(rNext)) {	
		@\textbf{\emph{\fence}}@; // before CAS
		bool res = CAS(&(right.next), rNext,mark(rNext));
		@\textbf{\emph{\flush}}@ (&right.next) ;
		@\textbf{\emph{\fence}}@; // before CAS/return
		if (res) {
			CAS(&(left.next), right, rNext));
			@\textbf{\emph{\flush}}@(&left.next) & @\textbf{\emph{\fence}}@;
			return false,true;
		}  
	}
	return true, false; // retry 
} 		 
\end{lstlisting}
\label{alg:harris}
\vspace{-0.7cm}
\end{figure}

\begin{figure}
		\vspace{-0.7cm}
	\caption{HLL Persistent Traverse and Find}
\begin{lstlisting}[frame=single, language=c++]	
bool, bool findCritical(List<@$\textbf{Node*}$@> nodes, @$\textbf{T}$@ key) @\label{line:startfind}@
	@$\textbf{Node*}$@ right = nodes.back();
	@\textbf{\emph{\fence}}@; // before return
	if (right.key != key) {	// no flush - immutable 
		return false, false;  // no key
	}
	return false, true; // key exists 
} 

List<@$\textbf{Node*}$@>,List<@$\textbf{Node*}$@> traverse(@$\textbf{Node*}$@ head, @$\textbf{T}$@ k) @\label{line:starttraverse}@
	List<@$\textbf{Node*}$@> parent, nodes;
	@$\textbf{Node*}$@ leftParent, left, right;
	while (true) {
		leftParent, left, right = head, head, null;
		nodes.clear();
		@$\textbf{Node*}$@ pred, curr = head, head; 
		@$\textbf{Node*}$@ succ = curr.next();
		while (isMarked(succ) || (curr.key < k)) { @\label{line:startinlooptraverse}@
			if (!isMarked (succ)) {
				nodes.clear();
				leftParent = pred; 
				left = curr;
				nodes.append (left); // found left node  @\label{line:applefttraverse}@
			} else {
				nodes.append (curr); 
			}
			pred = curr;
			curr = succ;
			if (!curr) 
				break;
			succ = curr.next(); 
		} @\label{line:endinlooptraverse}@
		right = curr; // found right node
		nodes.append (right); @\label{line:apprighttraverse}@
		if (right && isMarked (right.next)) { @\label{line:rechecktraverse}@
			continue;
		} else { 
			parent.append (leftParent); @\label{line:parenttraverse}@
			return parent, nodes; 
		} 
	}  
} @\label{line:endtraverse}@

bool deleteMarkedNodes(List <@$\textbf{Node*}$@> nodes) { @\label{line:startdeletemarked}@
	if (nodes.size() == 2) {
		@\textbf{\emph{\fence}}@; // before return
		return false; 
	}
	@$\textbf{Node*}$@ left, right = nodes.front(),nodes.back();
	@$\textbf{Node*}$@ leftNext = nodes[1];
	@\textbf{\emph{\fence}}@; //before CAS
	bool res = 	CAS(&(left.next), leftNext, right); @\label{line:casdeletemarked}@
	@\textbf{\emph{\flush}}@ (&left.next);
	if (res) {
		if (right && isMarked(right.next)) { @\label{line:rightokdeletemarked}@
			@\textbf{\emph{\flush}}@(&right.next) & @\textbf{\emph{\fence}}@; // before return @\label{line:flushdeletemarked}@
			return false; 
		} 
		@\textbf{\emph{\fence}}@; // before return
		return true; 
	}  
	@\textbf{\emph{\fence}}@; // before return
	return false; 
} @\label{line:enddeletemarked}@
\end{lstlisting}
\label{alg:harrisSearch}
\vspace{-0.7cm}
\end{figure}

	\renewcommand{\figurename}{Figure}

	\vspace{-0.3cm}
\section{Experimental Evaluation}\label{sec:exp}

We implement five \ourDS{s}: an ordered linked-list using the algorithm of \citet{harris2001pragmatic}, two binary search trees (BST) based on the algorithm of \citet{ellen2010non} and \citet{aravind14bst}, a hash table implemented by \citet{david2018logfree} based on Harris's linked-list, and a skiplist based on the algorithm of Michael~\cite{michael2002safe}. We compare the performance of the original, non-durable version of the algorithms to four ways of making it durable: our \durableOurDS{} (\emph{Traverse}), \citet{izraelevitz2016linearizability}'s construction (\emph{Izraelevitz}), the implementation of \citet{david2018logfree} (\emph{Log Free}) and Ramalhete et al.'s implementation for durable transactions \cite{rcaec19onefile} (\emph{Onefile}).

\vspace{-0.2cm}
\subsection{Setup}
We run our experiments on two machines; one with two Xeon Gold 6252 processors (24 cores, 3.7GHz max frequency, 33MB L3 cache, with 2-way hyperthreading), and the other with 64-cores, featuring 4 AMD Opteron(TM) 6376 2.3GHz processors, each with 16 cores. 

The first machine has 375GB of DRAM and 3TB of NVRAM (Intel Optane DC memory), organized as 12 $\times$ 256GB  DIMMS (6 per processor).
The processors are based on the new Cascade Lake SP microarchitecture, which supports the \texttt{clwb} instruction for flushing cache lines. We fence by using the \texttt{sfence} instruction.
We use \emph{libvmmalloc} from the PMDK library to place all dynamically allocated objects in NVRAM, which is configured to app-direct mode. All other objects are stored in RAM.
The operating system is Fedora 27 (Server Edition), and the code was written in c++ and compiled using g++ (GCC) version~7.3.1.

The second machine has 128GB RAM, an L1 cache of 16KB per core, an L2 cache of 2MB for every two cores, and an L3 cache of 6MB per half a processor (8 cores). The operating system is Ubuntu 14.04 (kernel version 3.16.0). \texttt{clwb} is not supported, so we used the synchronized clflush instruction instead. 
The code was written in C++ 11 and compiled using g++ version 8.3.0 with -O3 optimization.


We found that \citet{david2018logfree}'s code could not run on this NVRAM architecture for executions using more than 3 threads, so we only show comparisons to David et al's log-free data structures on the AMD machine.

On the NVRAM machine, we avoid crossing NUMA-node boundaries, since unexpected effects have been observed when allocating across NUMA nodes on the NVRAM .
Hyperthreading is used for experiments with more than 24 threads on the NVRAM machine. No hyperthreading is used on the DRAM machine. 
All experiments were run for 5 seconds and an average of 10 runs is reported.

On all the data structures, we use a uniform random key from a range $0,..,r-1$. We start by prefilling the data structure with $r/2$ keys. Keys and values are both 8 bytes. Unless indicated otherwise, all experiments use an insert-delete-lookup percentage of 10-10-80. For all data structures, we measured different read distributions, covering workloads A, B and C of the standard YCSB~\cite{YCSB10}.

For volatile data structures, memory management was handled with ssmem ~\citep{David15}, that has an epoch-based garbage collection and an object-based memory allocator. Allocators are thread-local, causing threads to communicate very little. 
For the durable versions, we used a durable variant of the same memory management scheme~\citep{Zuriel19}. 
The code for these implementations is publicly available in GitHub at https://github.com/michalfman/NVTraverse.

\subsection{Results on NVRAM}\label{subsec:nvram}
We begin our evaluation by examining the performance of various data structure implementations on the NVRAM Intel machine. We first examine the NVTraverse version of Harris's linked-list~\citep{harris2001pragmatic}. 

\begin{figure*}[!h]
  \begin{tabular}{@{}c@{}c@{}c@{}c@{}c@{}c}
      (a)&\includegraphics[width=0.64\columnwidth]{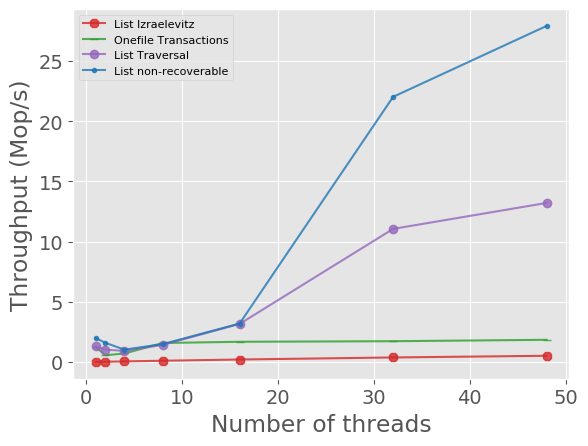}
      (b)&\includegraphics[width=0.64\columnwidth]{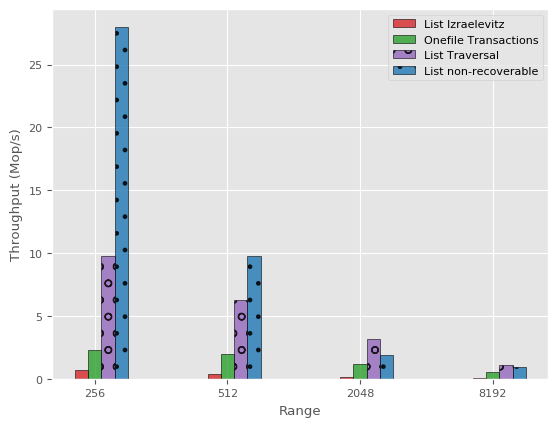} 
      (c)&\includegraphics[width=0.64\columnwidth]{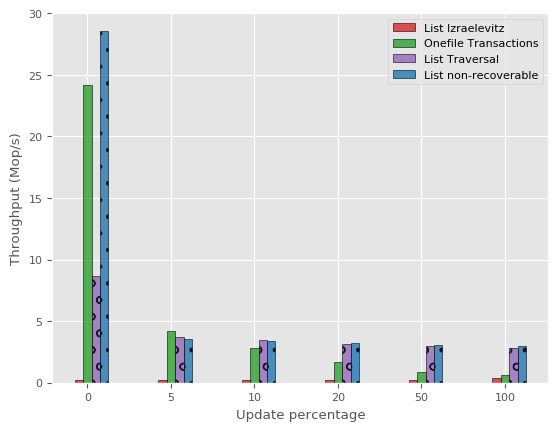} \\
      (d)&\includegraphics[width=0.64\columnwidth]{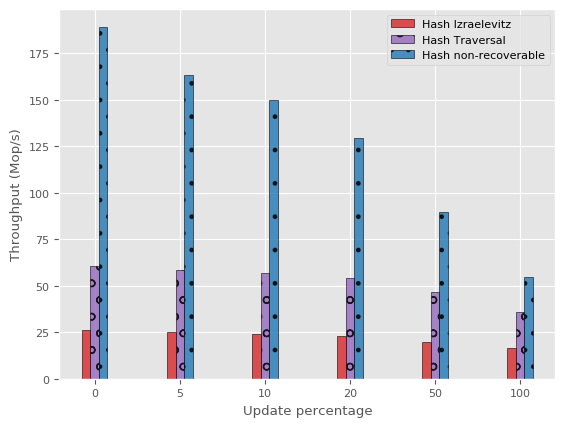} 
      (e)&\includegraphics[width=0.64\columnwidth]{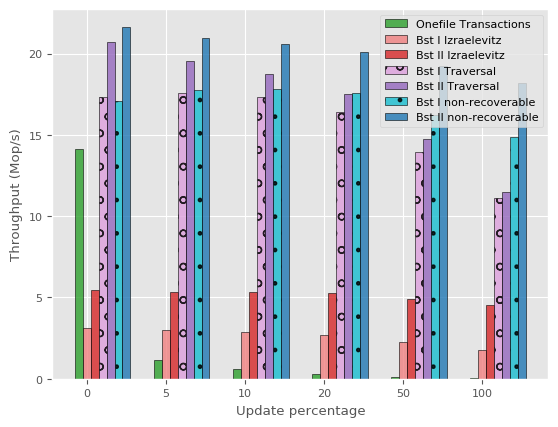}
      (f)&\includegraphics[width=0.64\columnwidth]{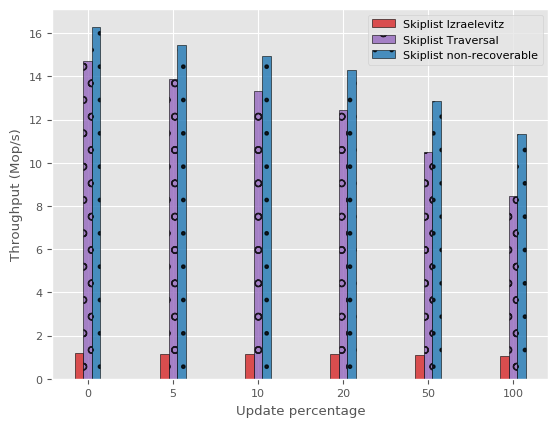}
  \end{tabular}\vspace{-.1in}
  \caption{\small NVRAM throughput results. (a). Linked-List, varying number of threads, 80\% lookups, 500 nodes. (b). Linked-List, varying size, 16 threads, 80\% lookups. (c). Linked-List, varying update percentage, 16 threads, 500 nodes. (d). Hash-Table, varying update percentage, 16 threads, 1M nodes. (e). BST, varying update percentage, 16 threads, 1M nodes. (f). Skip-List, varying update percentage, 16 threads, 1M nodes. }\label{fig:nvm}
  	\vspace{-.2in}
\end{figure*}

\textbf{List Scalability.}
We test the scalability as the number of threads increases, showing the results in \Cref{fig:nvm} (a). We initialize the list to have $512$ keys, and insert and delete keys within a range of $1024$. 

We note that while the non-durable version of the list outperforms the \durableOurDS{} by $2.1\times$, the latter outperforms \citet{izraelevitz2016linearizability}'s construction by $25.4\times$ and OneFile by $7.3\times$ on $48$ threads. While OneFile performs better than \citet{izraelevitz2016linearizability}'s construction, they scale similarly. The dramatic differences between \durableOurDS{s} and the other approaches hold true throughout all of the experiments that we've tried, highlighting the significant advantage of our approach.

The non-durable version and the \durableOurDS{} have a similar throughput up to 16 threads. However, the non-durable version scales better than the \durableOurDS.
Note that as the thread count increases, there are more flushes in the \durableOurDS{}, since each thread flushes a constant number of nodes. Each flush invalidates that cache line, meaning that as the number of threads increases, and cache misses become more likely. 

\textbf{List Size.}
Next we test how lists of different sizes affect performance. We expect that the size of the data structure may have a significant effect because a larger data structure means operations spend a larger fraction of their time traversing.
The results are shown in \Cref{fig:nvm} (b).
The first thing to note is that the original, non-durable version of the list greatly outperforms the \durableOurDS{} version for smaller lists. The non-durable version is better by $2.9\times$ for a short list of 128 nodes and by $1.5\times$ for the size of 256 nodes. However, the difference becomes less pronounced, and even inverts, as the list grows. This phenomenon can be explained by considering the traversal phase of operations on the list, and their role in determining the performance of a given implementation.
Recall that the \durableOurDS{} construction only executes a constant number of flushes and fences per traversal, and the non-durable version never executes flushes or fences. As the size of the data structures increases, the cost of the traversal outweighs the cost of persistence.
For the original list, the traversal is the primary source of delay, so the effect of increasing the traversal length in this implementation is starker than the effect of the same phenomenon in the durable version, which also spends significant time persisting. For the durable competitors, we observe the same trend as we saw in the list scalability test. The \durableOurDS{} construction outperforms \citet{izraelevitz2016linearizability}'s construction by $13.5\times$-$39.6\times$ and OneFile by $4.25\times$-$2.2\times$ on a range of 256-8192 respectively.

\textbf{List Update Percentage.}
We now consider the effect of varying percentage of updates (\Cref{fig:nvm} (c)). We note that these tests were run on a relatively small list (500 nodes), so the non-durable version outperforms the \durableOurDS{}.
Interestingly, the non-durable list sharply drops in throughput between $0\%$ updates and $5\%$ updates whereas the \durableOurDS{} version stays relatively stable across all update percentages. 
Since the list is less than 12Kb, it is small enough to fit in L1 cache, so there are virtually no cache misses on read-only workloads on the non-durable list. The \durableOurDS{} version still experiences cache misses because lookup operations perform \texttt{clwb}, invalidating the cache line. In our experience, it seems that in the current architecture, \texttt{clwb} and \texttt{clflushopt} yield the same throughput. In future architectures, we believe \texttt{clwb} will no longer invalidate cache lines and will perform better.

OneFile does extremely well in read-only workloads. This is because OneFile is optimized for such workloads.


\textbf{BST, Hash Table and Skiplist.}
We study how the Hash Table, BSTs and Skiplist behave under different YCSB-like workloads. The results are shown in Figures~\ref{fig:nvm} (d), (e), and (f) respectively. We only show OneFile's performance in the BST, since the patterns seen on OneFile are similar in all cases, and similar to the list.
We implemented two versions of the BST; one based on \citet{aravind14bst}'s tree, and the other on \citet{ellen2010non}'s tree. We saw that the amount of flushes and fences \citet{ellen2010non}'s tree executes is \emph{less} than in \citet{aravind14bst}. However, \citet{ellen2010non} performs worse than \citet{aravind14bst} in their volatile version, and the gap remains in the durable version.

We see that in the hash table, the non-recoverable version degrades twice as fast as the \durableOurDS{} as the number of updates grows. This is because allocating and writing nodes is more expensive than just reading. However, in the \durableOurDS{}, these costs do not form a bottleneck, because of the additional flush and fence instructions.
Interestingly, the skiplist and BSTs do not exhibit this behavior; in fact, the \durableOurDS{} version degrades faster than the non-recoverable version as the update percentage increases. This can happen due to the fact that as the number of updates increases, the likelihood of failed CASes increases, which is more meaningful than in the hash table. For the \durableOurDS{}, this means executing extra flush and fence instructions, which slows it down more in comparison to the non-recoverable version.

\begin{figure*}[!h]
	\begin{tabular}{@{}c@{}c@{}c@{}c@{}c@{}c}
		(g)&\includegraphics[width=0.64\columnwidth]{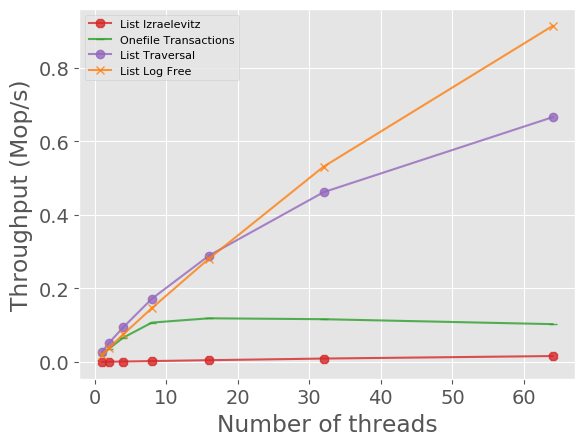} 
		(h)&\includegraphics[width=0.64\columnwidth]{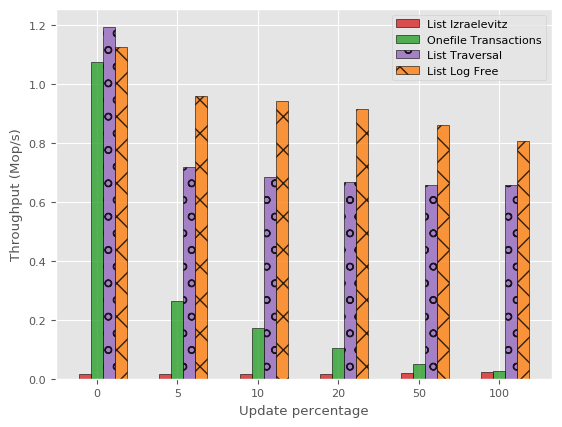}
		(i)&\includegraphics[width=0.64\columnwidth]{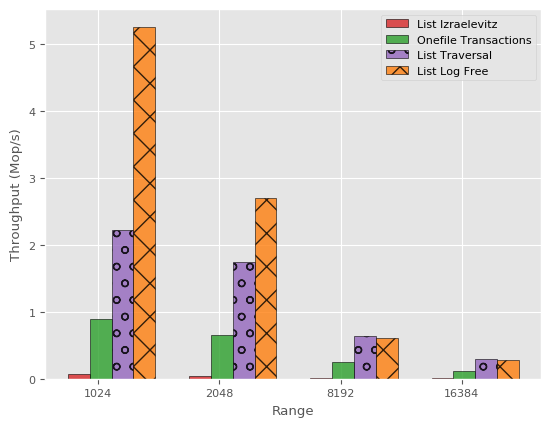}		\\
		(j)&\includegraphics[width=0.64\columnwidth]{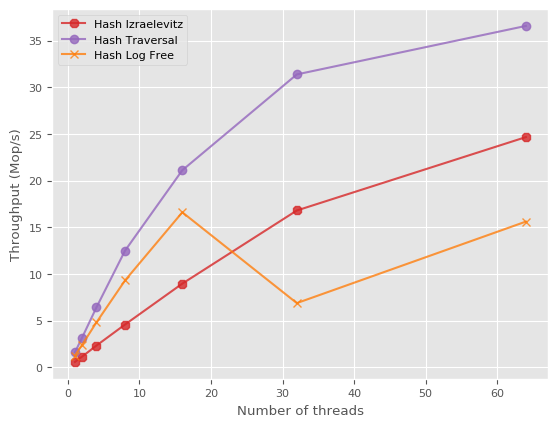}
		(k)&\includegraphics[width=0.64\columnwidth]{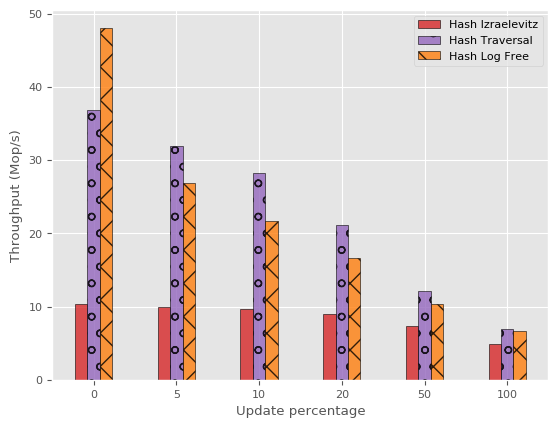}
		(l)&\includegraphics[width=0.64\columnwidth]{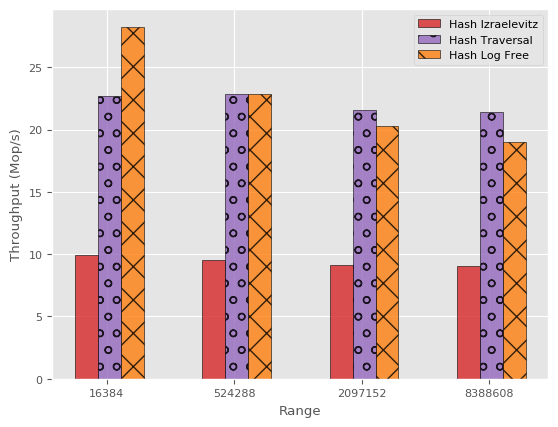}	\\
		(m)&\includegraphics[width=0.64\columnwidth]{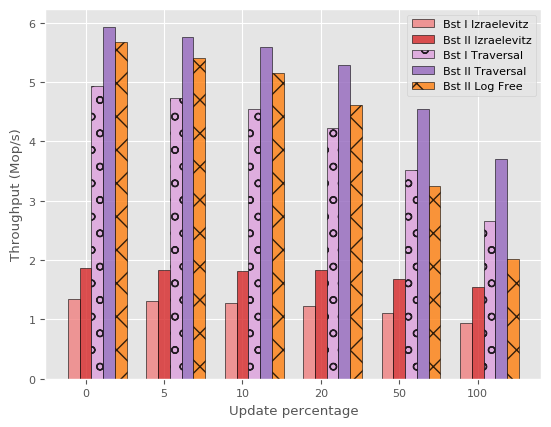}	
		(n)&\includegraphics[width=0.64\columnwidth]{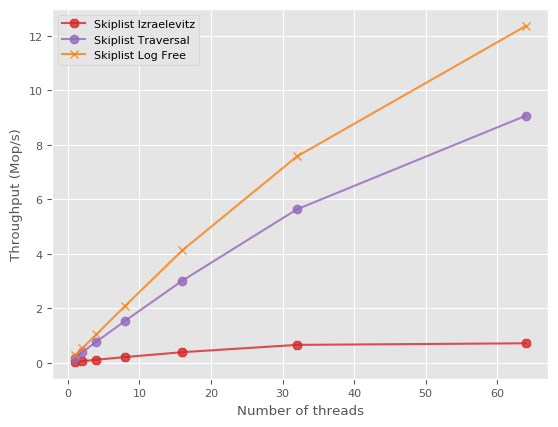}		
		(o)&\includegraphics[width=0.64\columnwidth]{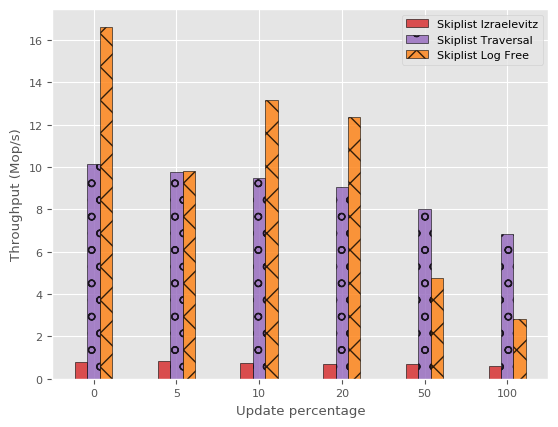}
	\end{tabular}\vspace{-.1in}
	  \caption{\small DRAM throughput results. (g). Linked-List, varying number of threads, 80\% lookups, 8000 nodes. (h). Linked-List, varying  update percentage, 64 threads, 8000 nodes. (i). Linked-List, varying size, 16 threads, 80\% lookups.  (j). Hash-Table, varying number of threads, 80\% lookups, 8M nodes. (k). Hash-Table, varying update percentage, 16 threads, 8M nodes. (l). Hash-Table, varying size, 16 threads, 80\% lookups. (m). BST, varying update percentage, 16 threads, 8M nodes. (n). Skip-List, varying number of threads, 80\% lookups, 8M nodes. (o). Skip-List, varying update percentage, 64 threads, 8M nodes.}\label{fig:dram}
	\vspace{-.2in}
\end{figure*}

\vspace{-0.2cm}
\subsection{Results on DRAM}\label{subsec:dram}
\vspace{-0.1cm}
We ran experiments on a machine with classic DRAM to compare with the algorithms of \citet{david2018logfree}. We ran \citet{david2018logfree}'s algorithms in the \emph{link-and-persist} mode. Link-and-persist, suggested by \citet{david2018logfree} and \citet{wang2018markbit}, is an optimization that allows avoiding flushing clean cache lines by tagging flushed words, but is not completely general, so we did not apply it to the \durableOurDS{} constructions. \citet{david2018logfree} actually present two optimizations in their paper, but the second one they present, called the link-cache, does not provide durable linearizability~\cite{izraelevitz2016linearizability}. At least one optimization must be selected.
%

\textbf{List.}
We ran with an initial size of 8192 nodes in the list with a range of 16,384  keys, varying the thread count. The results are shown in \Cref{fig:dram} (g). We notice that the linked-list algorithm of \citet{david2018logfree} outperforms ours by 15\%-50\%, from 32 to 64 thread counts; this is due to the fact that the link-and-persist technique reduces the number of flushes. With more threads, it is more likely that two threads get the same key, meaning that only one of them will have to flush.
 On the other hand, the \durableOurDS{} outperforms \citet{david2018logfree} by 40\%-16\%, for thread counts of 1-8. We believe this happens because of the same optimization. In link-and-persist, there is an extra CAS for each flush executed (to tag the word), and on a lower thread count, this optimization is less beneficial. So, our automatic construction does have a cost, but this cost is much smaller than \citet{izraelevitz2016linearizability}'s (by up to $ 56\times $).
 
We now test various update percentages, with 64 threads and same list size as above (\Cref{fig:dram} (h)). As we already noticed, the linked-list algorithm of \citet{david2018logfree} outperforms the  \durableOurDS{} by at most 1.37x at 20\% updates, as the flush-and-persist technique avoids some of the flushes. When there are read-only operations, our list is faster by 1.7x, again, as \citet{david2018logfree} executes some CASes to avoid extra flushes. We see the same trend in \Cref{fig:dram} (i), that shows 64 threads with varying key ranges. The bigger the list, the smaller the advantage of link-and-persist. OneFile \cite{rcaec19onefile} performs worse than our construction, as expected, by 1.1x-25x for 0\%-100\% update percentages respectively (\Cref{fig:dram}~(h)).

\textbf{Hash table.}
We observe the opposite trend in the hash table. In \Cref{fig:dram} (j), (k) and (l) we can see the scalability, various update percentages and ranges of keys respectively. The first two are filled with 8M nodes. For a fair comparison, due to the anomaly that we observe in \Cref{fig:dram} (j) on 32 threads, the update percentages are shown on 16 threads. 
With 0\% updates, the algorithm of \citet{david2018logfree} outperforms the \durableOurDS{}. This is because, to calculate the  bucket in the hash table, \citet{david2018logfree} use a bit-mask, assuming the table size is a power of 2. This is faster than \texttt{modulo}, a more general function that we use. In all the other comparisons, the \durableOurDS{} outperforms \citet{david2018logfree} by up to 30\% on 16 threads and 230\% on 64 threads. In a hash table with 8M nodes, the contention on every bucket is low, meaning that the price of the link-and-persist outweighs the benefit. This is clearer in \Cref{fig:dram} (l), which shows various range keys with 16 threads and 20\% update percentage. 

\textbf{Binary Search Tree.}
\Cref{fig:dram} (m) shows the results of various update percentages of the BST with 8M initial size and 16 running threads. We compare our two NVTraverse BST implementations with the implementation of \citet{david2018logfree}, as well as the two BST versions for \citet{izraelevitz2016linearizability}'s algorithm. 
\citet{david2018logfree} implements the durable version of \citet{aravind14bst} as well. As the contention is low, same as in the hash table, the CASes for marking the flushed nodes downgrades the performance of the same BST implementation by 4\%-83\% for 0\%-100\% of updates. 

\textbf{Skiplist.}
The scalability and varying update percentages of the skiplist are depicted in \Cref{fig:dram} (n) and (o). \Cref{fig:dram} (n) shows the scalability for an initial size of 8M nodes and 20\% updates. As it executes one less flush in every search operation, the implementation of \citet{david2018logfree} performs better than the \durableOurDS{} in a read dominated workload. For 64 threads and 20\% updates, \citet{david2018logfree} is 1.3x better, and it reaches the maximum difference of 1.63x in 0\% of updates. However, as seen in 
\Cref{fig:dram} (o), as the workload becomes more write dominant, the performance degrades; it benefits less from the flush that was saved in the search operation. The \durableOurDS{} gets better throughput by 1.68x and 2.4x on 50\% and 100\% updates.


\subsection{Other Architectures}
\Changed{We showed the evaluation of our transformation on two different architectures. We believe that our persistent transformation is hardware-agnostic and relevant to other frameworks that satisfy other memory models. However, the instructions that are used for executing flushes and fences should be adjusted accordingly. For instance, in an ARM architecture, a flush instruction may translated to the \emph{DC CVAP} instruction and a fence instruction may be executed by calling to a full \emph{DSB} instruction~\cite{arm,raad2019model}.}
	\vspace{-0.1cm}
\section{Related Work}

%
%

NVRAM has garnered a lot of attention in the last decade, as its byte-addressability and low latency offer an exciting alternative to traditional persistent storage.
Several papers addressed implementing data structures for file systems on non-volatile main memory~\cite{chen2015persistent,lee2017wort,lejsek2009nv,xu2016nova,venkataraman2011consistent,yang2015nv}. 


Recipe~\cite{lee2019recipe} provides a principled approach to making some index data structures persistent. While on the surface, their contribution is similar to ours, their original approach does not always yield correct persistent algorithms, even for data structures that fit their prescribed conditions. The ArXiv version of their paper has updated Conditions 1 and 2 to account for this. The new formal conditions require flush and fence instructions for every read and write, similarly to the requirement of Izraelevitz et al.~\cite{izraelevitz2016linearizability}. While they note that in some situations, one can leave out some of these flushes and fences, they leave it to the user to hand-tune their algorithms to do so. Recipe thus exemplifies the difficulty of providing a correct, general and efficient solution, and the importance of having one. 
In our work, we provide a general and automatic way to reduce the required flush and fence instructions for traversal data structures that yields efficient persistent data structures, and we provide a proof that this transformation is correct.

Mnemosyne~\citep{volos2011mnemosyne} provides a clean programming interface for using persistent memory, through \emph{persistent regions}. 
Atlas~\citep{chakrabarti2014atlas} provides durability guarantees for general lock-based programs, but does not capture lock-free algorithms.
Some general constructions for persistent algorithms have been proposed. 
\citet{cohen2018inherent} present a universal construction. 
%
\citet{izraelevitz2016linearizability} presented a general construction that converts a given linearizable algorithm into one that is durably linearizable,
but 
they do not capture semantic dependencies like those captured by our \ourDS.
Another approach for general constructions for persistent concurrent algorithms uses transactional memory.
This involves creating persistent logs to either undo or redo partial transactions~\citep{cohen2017efficient,correia2018romulus,liu2018ido,wang2014scalable}. 
While these approaches are general, they suffer from the usual performance setbacks associated with transactional memory.

\citet{friedman2018persistent} presented a hand-tuned implementation of a durable lock-free queue, based on the queue of \citet{michael1996simple}, and presented informal \emph{guidelines} for converting linearizable data structures into durable ones. 
Based on these guidelines, \citet{david2018logfree} implement several durable data structures. 
\citet{david2018logfree} achieve this by carefully understanding each data structure to find its dependencies, and only intuitively argue about correctness. Our definition of \ourDS{s} formalizes some dependencies in a large class of algorithms 
and removes the need for expert understanding of persistence and concurrency.

Other general classes of lock-free algorithms have been defined. 
\citet{brown2014general} defined a general technique for lock-free trees, and
\citet{timnat2014practical} defined \emph{normalized} data structures. These classes were defined with different goals in mind and do not aid in finding dependencies that are critical for efficient persistence.

Another line of work focuses on formally defining persistency semantics for different architectures~\cite{raad18oopsla, raad20popl}.

\vspace{-0.15cm}

	\vspace{-0.3cm}
\section{Conclusions}

Recent NVRAM offers the opportunity to make programs resilient to power failures. However this requires that the persistent part of memory is kept in a consistent and recoverable state. In general this can be very expensive since it can require flushing and fencing between every read or write. This renders caching almost useless.  In this paper we considered a broad class of linearizable concurrent algorithms that spend much of their time traversing a data structure before doing a relatively small update. The goal is to avoid any flushes and memory barriers during the traversal (read-only portion). For a balanced tree, for example, this can mean traversing $O(\log n)$ nodes without flushing, followed by $O(1)$ flushes and fences. We describe conditions under which this is safe. Although the conditions require some formalism, we believe that in practice they are quite natural and true for many if not most concurrent algorithms. We studied several algorithms under this framework and experimentally compared their performance to state-of-the-art competitors. We run the experiments on the recently available Intel Optane DC NVRAM.  The experiments show a significant performance improvement using our approach, even beating hand-tuned algorithms on many workloads.

	\vspace{-0.4cm}
\begin{acks}
	
We thank Michael Bond for his helpful comments on this paper. This work is supported by the United States - Israel BSF grant No. 2018655, by NSF grants CCF-1910030 and CCF-1919223, by an Azrieli PhD Fellowship, a Microsoft PhD Fellowship, and an NSERC PGSD Scholarship.
	
\end{acks}


\bibliographystyle{plainnat}
\bibliography{strings,biblio}

\pagebreak
\end{document}